\documentclass[useAMS,usenatbib]{mn2e}
\usepackage{graphicx,amssym}
\def\oii{[{O~\sc ii}]}
\def\neiii{[{Ne~\sc iii}]}
\def\oiii{[{O~\sc iii}]}
\def\nii{[{N~\sc ii}]}

\def\hi{{H~\sc i}}
\def\hei{{He~\sc i}}
\def\heii{{He~\sc ii}}

\newcommand{\bc}{\begin{center}}
\newcommand{\ec}{\end{center}}
\newcommand{\cf}{\ifmmode C_f\else $C_f$\fi}

\title[Evolution of the Most Massive Galaxies to $z$=0.6]
       {Evolution of the Most Massive Galaxies
 to $z$=0.6: I. A New Method for Physical Parameter Estimation}

\author[Chen et al.]{
\parbox[t]{\textwidth}{\raggedright
Yan-Mei Chen$^{1,2}$\thanks{Email: chen@astro.wisc.edu},
Guinevere Kauffmann$^3$,
Christy A. Tremonti$^1$,
Simon White$^3$,
Timothy M. Heckman$^4$,
Katarina Kova\v{c}$^3$,
Kevin Bundy$^5$,
John Chisholm$^1$,
Claudia Maraston$^6$,
Donald P. Schneider$^7$,
Adam S. Bolton$^8$,
Benjamin A. Weaver$^9$,
Jon Brinkmann$^{10}$
}\\
\vspace*{6pt}\\
$^1$Department of Astronomy, University of Wisconsin-Madison, 1150 University Ave, Madison, WI 53706, USA\\ 
$^2$Department of Astronomy, Nanjing University, Nanjing 210093, China\\ 
    Key Laboratory of Modern Astronomy and  Astrophysics (Nanjing University), Ministry of Education, Nanjing 210093, China\\
$^3$Max--Planck--Institut f\"ur Astrophysik,
    Karl--Schwarzschild--Str. 1, D-85748 Garching, Germany\\
$^4$Department of Physics and Astronomy, The Johns Hopkins University, 3400 North Charles Street, Baltimore, MD 21218\\
$^5$Astronomy Department, University of California, Berkeley, CA 94705, USA\\
$^6$Institute of Cosmology and Gravitation, University of Portsmouth, Dennis Sciama Building, Burnaby Road, Portsmouth PO1 3FX\\
$^7$Department of Astronomy and Astrophysics, The Pennsylvania State University, 525 Davey Laboratory, University Park, PA 16802, USA\\
$^8$Department of Physics and Astronomy, University of Utah, Salt Lake City, UT 84112\\
$^9$Center for Cosmology and Particle Physics, New York University, New York, NY 10003 USA\\
$^{10}$Apache Point Observatory}        
\begin{document}



\maketitle

\label{firstpage}
\begin{abstract}
We use principal component analysis (PCA) to estimate stellar masses,
mean stellar ages, star formation histories (SFHs), dust extinctions
and stellar velocity dispersions for a set of $\sim 290,000$ galaxies
with stellar masses greater than $10^{11} M_{\odot}$ and redshifts in
the range $0.4<z<0.7$ from the Baryon Oscillation Spectroscopic Survey
(BOSS). 
We find  that the fraction of  galaxies
with active star formation first declines with increasing stellar mass,
but then {\em flattens} above a stellar mass of
$10^{11.5} M_{\odot}$ at $z\sim 0.6$. This is in striking contrast
to $z \sim 0.1$, where the fraction of galaxies with
active star formation declines monotonically with stellar mass.   
At  stellar masses of $10^{12} M_{\odot}$, therefore, the evolution in the fraction of 
star-forming galaxies from $z \sim 0.6$ to the present-day  
reaches a factor of $\sim 10$. 
When we stack the
spectra of the most massive, star-forming galaxies at $z \sim 0.6$, we
find that half of their \oiii\ emission is produced by AGNs. The black
holes in these galaxies are accreting on average at $\sim$ 0.01 the
Eddington rate.  To obtain these results, we use the stellar
population synthesis models of \citet{bc03} to generate a
library of model spectra with a broad range of SFHs, metallicities,
dust extinctions and stellar velocity dispersions.  The PCA is run on
this library to identify its principal components over the rest-frame
wavelength range $3700 - 5500$\AA.  We demonstrate that linear
combinations of these components can recover information equivalent to
traditional spectral indices such as the 4000\AA\ break strength and
H$\delta_A$, with greatly improved signal-to-noise ratio. In addition, the
method is able to recover physical parameters such as stellar
mass-to-light ratio, mean stellar age, velocity dispersion and dust
extinction from the relatively low S/N BOSS spectra.  We examine in
detail the sensitivity of our stellar mass estimates to the input
parameters in our model library, showing that almost all changes
result in systematic differences in log $M_*$ of 0.1 dex or less. The
biggest differences are obtained when using different population
synthesis models -- stellar masses derived using \citet{maraston11} 
models are systematically smaller by up to 0.12 dex at young
ages.
\end{abstract}


\section{Introduction}
\label{sec:intro}
The mass of a galaxy is perhaps its most fundamental physical property.
Many galaxy properties, such as 
metallicity \citep{tremonti04} 
and recent star formation history \citep{salim05, juneau05, heavens04, kauffmann03a}, exhibit tight
correlations with stellar mass.
Others, such as the $\alpha-$enhancement \citep{thomas04, thomas05, 
gallazzi06} correlate best with
stellar velocity dispersion, which is a measure of the mass contained in
the bulge-dominated region of the galaxy. The baryonic mass of a galaxy
is correlated with the mass of the dark matter halo
in which most of its stars formed \citep{kauffmann97, benson00, 
kauffmann99, pearce01, wangl06}. 
Finally, studies of how the scaling between galaxy mass and other physical
properties evolve with redshift place important constraints on 
the mass assembly histories of galaxies, as well as on the processes
that regulate  star formation and feedback in these systems 
\citep{chen10, wang07}.

Galaxy masses are traditionally estimated in two ways: (i) from the motions
of stars and/or gas. This method measures the total mass contained interior
to the radius where one measures these motions, and will include not only
the the directly observed material, but also the dark matter present
in the galaxy; (ii) from estimates of stellar mass-to-light ($M_*/L$) ratio 
based on  fits of multi-band photometry to a grid of composite 
stellar population models. In the last decade, the efficiency with 
which multi-band photometry has been collected 
from ground- and space-based observatories has greatly increased, 
enabling  stellar mass-to-light ratios to be estimated for large samples
of galaxies. If spectra are available, 
the information carried by stellar absorption lines enables
the recent star formation history of a galaxy to be more accurately
determined \citep{kauffmann03a, gallazzi09, panter04}. Constraints on $M_*/L$ are also    
improved with spectroscopic information, but in practice the stellar masses estimated from  
multi-band photometry and spectral indices for low-redshift galaxies with
luminosities $\sim L_*$  are consistent 
with less than $\sim$0.1 dex scatter \citep[e.g.,][]{bell03, drory04, 
salim05, borch06}. The stellar masses so obtained correlate strongly with dynamically-measured 
masses  \citep[e.g.,][]{bell01, van05, cappellari06}. 

The agreement between different methods for 
estimating stellar mass has led to a certain amount of complacency in the
community (see Conroy et al. 2009 for a review). 
It is important to be
aware of the following: 
(1)Uncertainties in the inputs to the stellar population synthesis codes  
used to generate the model galaxy grids are a source of systematic 
error in stellar mass estimation. \citet{maraston05} has reported that
thermally pulsing asymptotic giant branch (TP-AGB) stars contribute 
significantly to the near-infrared light of galaxies. Because
the physics driving the pulsations is poorly understood, this phase
of stellar evolution may not be very well represented in many current models. 
(2) At high redshifts ($z\sim1$), one often lacks access to  rest-frame
near-infrared data, so star formation histories must be estimated using the
shape of the spectral energy distribution (SED) in the UV/optical. This limitation can lead
to systematic offsets between the stellar masses derived for  
galaxy samples with different redshifts, even if the
true SEDs are the same  \citep{kannappan07}. (3) If the model galaxies
do not provide a correct representation of the star formation histories
of the galaxies in the sample or of the transmission of the 
starlight to the observer, this will also lead to errors; for example, 
dusty galaxies and galaxies that have experienced recent starbursts  
have  poorly-measured masses if the model library includes only    
galaxies with smooth star formation histories and no dust.
(4) Robust stellar masses cannot be estimated if the S/N of the data is poor.
This last point is perhaps an obvious one, but it is important to remember that
quoted errors on stellar mass estimates are as important as the
estimates themselves.

In this paper, we present a method based on a principal component
analysis \citep[PCA;][]{madgwick03a, madgwick03b, lu06}. to estimate galaxy physical parameters
from rest-frame optical galaxy spectra
using stellar population synthesis models. 
These parameters include  stellar masses, 
metallicities, dust extinction, velocity dispersions, and
estimates of the recent star formation histories of galaxies such as
luminosity-weighted ages and the recent fraction of the stellar mass formed in bursts.
Unlike much previous work, which relied on narrow-band indices defined
in the vicinity of a limited set of stellar absorption lines to estimate
these parameters, our method employs all the information contained in the
rest-frame wavelength range of the spectrum between 3700 and 5500\AA.
Because our method makes use of the full spectrum, it can be applied to
lower S/N data. In addition, the chosen wavelength range    
is accessible out to $z \sim 0.8$, even in optical spectra;
this means the analysis can be applied to both low- and high-redshift galaxy samples
in a consistent manner.
 
We apply our method to a set of spectra from the
Sloan Digital Sky Survey Data Release 7 \citep[DR7;][]{abazajian09},
as well as a new sample of 290,000  spectra of luminous galaxies 
from the Baryonic Oscillation Spectroscopic Survey \citep[BOSS;][]{eisenstein11, schlegel09}.
We present spectrum-based stellar mass estimates for these galaxies, as
well as their recent star formation histories, and use this information
to assess how the recent star formation histories of the most massive
galaxies in the Universe have evolved from $z \sim 0.55$ to the present day.

Our paper is arranged as follows. In \S2, we introduce the two data sets. 
Our methodology for estimating  stellar mass and recent SFH is 
developed in \S3. The improvements of this new method are 
discussed in \S4. We 
compare our stellar masses with those from multi-band photometry 
in \S5. We also discuss systematic effects in our estimates of   
stellar mass. 
Our results on the evolution of massive galaxies are presented in \S6. 
We use the cosmological parameters 
$H_0=70~{\rm km~s^{-1}~Mpc^{-1}}$, $\Omega_{\rm M}=0.3$ 
and $\Omega_{\Lambda}=0.7$ throughout this paper.

\section{The SDSS Data}
\subsection{SDSS DR7}
The Sloan Digital Sky Survey \citep[SDSS;][]{york00}  obtained photometry 
of nearly a quarter of the sky and spectra of about one 
million objects. A drift-scanning mosaic CCD camera \citep{gunn98} mounted on 
the SDSS 2.5m telescope at Apache Point Observatory \citep{gunn06}
imaged the sky in the $u, g, r, i, z$ bands \citep{fukugita96}. 
The imaging data are 
astrometrically  \citep{pier03} and  photometrically \citep{hogg01, pad08}
calibrated  and used to 
select stars, galaxies, and quasars for follow-up fiber 
spectroscopy. 

The seventh data release \citep{abazajian09} of the SDSS 
includes  $\sim$930,000 galaxy spectra. The spectra have a wavelength
coverage of $3800 - 9200$\AA\ and  are taken through 
$3^{\prime\prime}$ diameter fibers.
The instrumental resolution is $R \sim 2000$ and the dispersion
is $\Delta{\rm log} \lambda = 10^{-4}$, where $\lambda$ is 
the wavelength in Angstroms. 
The median S/N per pixel of the spectra ranges from 
4 to 30, with a median of  14. 
 In this paper, we make use of
the spectra of galaxies  from the
``Main Sample'' \citep{strauss02}, which have   
Petrosian magnitudes in the range $14.5 < r< 17.6$ after correction 
for foreground galactic extinction using the reddening maps 
of \citet{schlegel98}. The redshift range spanned by these galaxies is 0.01 to 0.30 (see Figure \ref{zdist}).

Stellar masses and star formation rate for the Main Sample have been publicly available
since 2008 in the MPA/JHU catalog\footnote{The MPA/JHU stellar mass catalog can be downloaded 
from http://www.mpa-garching.mpg.de/SDSS/DR7 and is 
available through the SDSS Catalog Archive Server as described in
\citet{aihara11}.}. The stellar masses are estimated from broad-band 
$u, g, r, i, z$ photometry 
(``model" fluxes are used, see Stoughton et al. (2002) for definitions of the magnitudes). 
The total fluxes are corrected for emission line contribution by assuming 
that the relative contribution of emission lines to the broad-band 
magnitudes is the same inside the fiber as outside. 
A large grid of model star formation histories is generated    
following the methodology described in  \citet{kauffmann03a},  and comparison of
the observed colors with predictions from these models allows one to derive  
maximum likelihood estimates of the 
$z$-band stellar mass-to-light ratios ($M_*/L_z$) of the galaxies. 
Stellar masses are computed  by multiplying 
$M_*/L_z$ by the  $z$-band ``model" luminosity $L_z$. Masses derived
from the 5-band photometry are consistent with 
previous stellar masses derived using  spectral features (D4000 \& H$\delta_A$ 
in the case of Kauffman et al. 2003a and a total of five indices in the case of Gallazzi et al. 2005) with an r.m.s scatter of 0.1 dex in log $M_*$. 
The SFRs are derived from nebular emission lines \citep{brinchmann04}.

The large wavelength coverage and high S/N of the
 DR7 spectra, and the fact that robust stellar mass estimates are publicly 
available, means that the DR7 main sample (hereafter we refer this sample as the DR7 sample for convenience) is very well-suited 
for developing and  testing 
our new method of physical parameter estimation (see \S5).

\subsection{Baryon Oscillation Spectroscopic Survey}
The SDSS-III project has completed an additional 3000 ${\rm deg}^2$ of imaging 
in the southern Galactic cap in a manner identical to the original SDSS imaging \citep{aihara11}.
BOSS is obtaining spectra of a selected subset of 
1.5 million luminous galaxies to  $z \sim 0.7$. 
(N. Padmanabhan et al. 2011, in preparation).  
The spectrographs have been  significantly upgraded from 
those used by SDSS-I/II, with improved CCDs with better red response, 
high throughput gratings, and an increased number of fibers (1000 
instead of 640).
The new fibers are  $2^{\prime\prime}$ in diameter and the spectra
cover the  wavelength range  3600$-$10,000\AA, 
at a resolution of about 2000. 
The BOSS galaxy spectra have median S/N per pixel ($\Delta{\rm log} \lambda = 10^{-4}$) of  $\sim$2.5. 

The details of how targets are selected from the SDSS photometry
will be described in detail in Padmanabhan et al.~(in preparation).
The sample we analyze here is the ``CMASS'' sample (so-named,
because it is very approximately stellar-mass limited). The sample
is defined using the following cuts:
\begin{equation}
\begin{array}{c}
  d_\perp>0.55 \quad{\rm and}\quad
17.5 < i < 19.9 \quad {\rm and}\quad i_{\rm fiber2} < 21.5 \\
  i < 19.86+1.6\,(d_\perp-0.8) \quad {\rm and} \quad r-i<2
\end{array}
\label{eqn:cuts}
\end{equation}
where $d_\perp$ is a ``rotated'' combination of colors defined as
$d_\perp=(r-i)-(g-r)/8$.
$i_{\rm fiber2}$ is the magnitude of the galaxy measured within
a $2^{\prime\prime}$ diameter aperture; this is the amount of light that enters the fiber.   
We note that all color cuts are defined using ``model" magnitudes,
whereas magnitude limits are given in terms of ``cmodel" magnitudes.
Two additional cuts are introduced  to reduce  contamination by stars;
$z_{\rm psf}-z_{\rm model}\ge 9.125-0.46\,z_{\rm model}$ and $i_{\rm psf}-i_{\rm model}> 0.2 + 0.2\times (20.0-i_{\rm model})$ (``psf'' refers to the
$psfMag$ quantity in the SDSS database).  Redshifts are successfully determined for 95.3\% of 
CMASS galaxies with no apparent dependence on galaxy type.

\begin{figure}
\bc
\hspace{-0.6cm}
\resizebox{8.5cm}{!}{\includegraphics{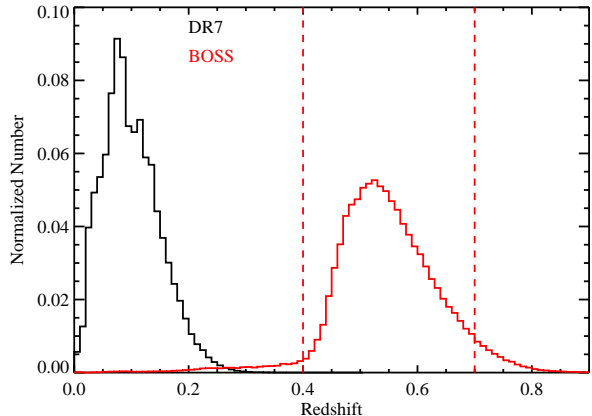}}\\%
\caption{Black: the redshift distribution of the DR7 sample; 
red:  the redshift distribution of the``CMASS"
sample of  BOSS high-$z$ luminous galaxies.  
The two vertical dashed lines indicate the redshift limits of the ``CMASS" sample used in our analysis.
\label{zdist}}
\ec
\end{figure}

These cuts are designed to select massive galaxies with $z > 0.45$.
The constraint that $d_\perp > 0.55$ identifies galaxies that lie at high
enough redshift so that the 4000 \AA\ break has shifted beyond the 
observer frame $r$-band, leading to red $r-i$ colors. The cut on the $i$-band magnitude  
of the galaxy is designed to produce a sample that is roughly complete down
to a limiting stellar mass of ${\rm log} M_* \sim 11.2$ (we will come back to this point later).  
The analysis presented in this paper includes the $\sim 290,000$  CMASS spectra
obtained prior to July 2011 which will be made publicly
available in SDSS Data Release 9.  
We restrict our analysis to galaxies in the redshift range 
$0.4<z<0.7$ (see Figure \ref{zdist}).  

\section{The method}

Because the CMASS galaxies are faint, the errors on the broad-band colors are
large; this is especially true of colors that include the $u$-band. The median magnitude 
errors are 0.62, 0.16, 0.06, 0.04, 0.09 for $u, g, r, i, z$-band measurements, respectively.
In this section, we describe a new method for estimating 
galaxy properties, including stellar masses and  recent SFHs, 
using the SDSS spectra rather than photometry. As we will show, this method
can be applied successfully to low S/N data, such as that from the BOSS survey. 

The main concept  underlying our approach is the fact that a
galaxy spectrum can be described as a number of orthogonal 
principal components (PCs). When one applies a  PCA decomposition to
model galaxy spectra, one finds that the PCs can be related to    
the physical properties of galaxies, such as their stellar mass-to-light ratios,  
their recent SFHs, the average age of their stellar populations, 
and their velocity dispersions. By decomposing each observed 
spectrum into the same set of PCs,  
and by comparing the amplitudes of the components with
those calculated for the model galaxies, we   
obtain the likelihood distribution of a given parameter $P$ 
in the space of the values of $P$ allowed by the models.   
In this work, we calculate likelihood distributions by comparing the amplitudes 
of the first seven PCs (\S3.4.1 clarifies this choice)  
to those calculated for  the galaxies in the model library. 
In the following sections, we provide a step-by-step description of our method.  

\subsection{The  spectral range used in the analysis}
We follow two criteria to select the rest-frame spectral range: (1) it 
should be as wide as possible so as to make optimum use of all the information
carried in the spectrum;  
(2) to minimize systematic effects, the same spectral range should be used
in the analysis of both the low redshift and the high redshift galaxies.  
Considering the redshift 
and wavelength coverage of DR7 ($0.005 < z < 0.3$, $3800 < \lambda < 9200$\AA) and 
of BOSS ($0.4 < z < 0.7$, $3600 < \lambda < 10,000$\AA) galaxy samples, we choose 
rest-frame 3700\AA\ to 5500\AA\ 
for the current analysis.  Spectral features located in this
region, including  the 4000\AA\ break strength 
(D4000) and Balmer absorption lines,  
provide important information about the stellar populations and 
recent SFHs of the galaxies.

Single stellar population (SSP) models that are well-matched in
spectral resolution (e.g. from Bruzual \& Charlot 2003, hereafter BC03     
and from Maraston \& Stromback 2011, hereafter M11) are also available 
over this wavelength range. BC03 has a spectral resolution of 75~${\rm km~s}^{-1}$, while 
M11 spectral resolution is 65~${\rm km~s}^{-1}$, 
similar to the instrumental resolution, $\sim75~{\rm km~s}^{-1}$, 
of the DR7 and BOSS spectra.

\subsection{Model library}
We generate a library containing 25,000 realizations\footnote{In order 
to access the effects of the ``resolution" of our model parameters, i.e. how estimation of physical parameters depends on 
the number of models,  we increase the model number to 100,000, finding that 
25,000 models are enough to converge. \citet{salim07} reached the same
conclusion.} of different SFHs using SSP models from BC03. The model library is parametrized as follows
\begin{enumerate}
\item {\em SFHs.} Each SFH consists of three parts: 
a) an underlying continuous model, b) a series of super-imposed  stochastic 
bursts, c) a random probability for star formation
to stop exponentially (i.e. truncation).
In the continuous model, stars are formed from 
the time $t_{\rm form}$ to the present according to the following functional form: 
${\rm SFR}(t) \propto {\rm exp}[-\gamma (t-t_{\rm form})]$.
The formation time $t_{\rm form}$ is uniformly distributed 
between 13.5 and 1.5 Gyr and the 
star formation inverse time-scale $\gamma$ is uniformly distributed
over the range 0$-$1 Gyr$^{-1}$.  
The main parameter that  controls the bursts is the amplitude $A$,  defined 
as the fraction of stellar mass produced during the burst 
relative to the total mass formed by the 
continuous model. $A$ is logarithmically distributed between 0.03 and 4. 
During the burst, stars are formed at a 
constant rate that is independent of $A$ for a time $t_{\rm burst}$,
which is uniformly distributed between $3 \times 10^7 - 3 \times 10^8$yr.
Bursts  occur at all times after  $t_{\rm form}$ with equal probability. 
The probability is set in such a way that 15\%
of the galaxies in the library experience a burst in the last 2 Gyr.

The existence of a population of massive, 
compact  ``post-starburst'' galaxies at high
redshifts with little or no ongoing star formation \citep[e.g.,][]{kriek06, kriek09}
has triggered us to add possible  
truncations to the SFHs. For 30\% galaxies in the library, we truncate the star formation at a random time
in the past. Following the truncation event,
the star formation rate evolves as  ${\rm SFR}(t>t_{\rm cut}) \sim {\rm SFR}(t_{\rm cut})~ {\rm exp}[-(t-t_{\rm cut})/\tau]$,
where $t_{\rm cut}$ is the truncation time and log$\tau$ is
uniformly distributed in the range  7 to 9. 
We note that in \citet{kauffmann03a}, the fraction of galaxies
with bursts in the last 2 Gyr  was set to be 50\%. We have 
reduced this fraction to 10\% (after truncations are included) 
because it provides a more uniform distribution in the light-weighted 
age of the models.
The influence of the choice of the fraction of galaxies
with recent bursts on our physical parameter estimation is discussed in \S5.3.

\item {\em Metallicity.} The BC03 model library includes six metallicities ranging from 
0.005 to 2.5$Z_\odot$; we interpolated the BC03 model grids in metallicity in log-space.
95\% of the  model galaxies in our library are distributed 
uniformly in metallicity from  $0.2 - 2.5 Z_\odot$;
5\% of the model galaxies are  distributed uniformly between
$0.02$ and  $0.2 Z_\odot$. The reason for including 
these very low metallicity models is that 
 \citet{maraston09} have found that the colors  predicted for
simple single stellar populations do not provide a good match to the observed
colors of the reddest galaxies in the SDSS.   
Adding a small fraction (typically 3\% of the total stellar mass) 
of old metal-poor stars allows one to achieve a significantly better match to the observations.
\item {\em Dust extinction.} Dust extinction is modelled
using  the two-component model described 
in \citet{charlot00}.  The  $V$-band optical depth
has a Gaussian distribution over the range
$0 < \tau_V < 6$, with a peak at 1.2 and 68\% of the total
probability distribution distributed over the range $0-2$. 
Our adopted  prior distribution of $\tau_V$ values is motivated by the observed
distribution of Balmer decrements in SDSS spectra \citep{brinchmann04}. 
The fraction of the optical depth that affects stellar populations 
older than $10^7$yr is parametrized as  $\mu$, which
is again modeled as a Gaussian with a peak at $\mu = 0.3$
and a 68 percentile range of $0. 1-1$.
\item {\em Velocity dispersion.}  Each of the model spectra is 
convolved to a velocity that is uniformly  distributed 
over the range of values from  50 to 400${\rm km~s}^{-1}$.
\end{enumerate}

We adopt the universal initial mass function (IMF) 
given in \citet{kroupa01}.  For each model in the library, 
we store the following properties:
\begin{enumerate}
\item the spectrum over the wavelength range ($91 - 160,000$\AA);
\item the strengths of the D4000 and H$\delta_A$ indices,
measured in the same way as in the SDSS spectra\footnote{The 
values of D4000 and H$\delta_A$ for  DR7 galaxies can be found 
in the MPA/JHU catalog. The bandpasses over 
which these two indices are measured are given in \citet{balogh99} and \citet{wortheyott97}.};
\item the $r$-band luminosity-weighted age, which is defined as 
$t_r = \int_0^t [d\tau~{\rm SFR}(t-\tau)~f_r(\tau)~\tau]/\int_0^t [d\tau~{\rm SFR}(t-\tau)~f_r(\tau)]$, where 
$f_r(\tau)$ is the total $r$-band flux produced by stars at age $\tau$;
\item the mass-weighted age, which is  calculated as 
$t_m = \int_0^t [d\tau~{\rm SFR}(t-\tau)~\tau]/\int_0^t [d\tau~{\rm SFR}(t-\tau)]$;
\item the $i$-band and $z$-band stellar mass-to-light ratios, 
$M_*/L_i$ and $M_*/L_z$, of the model at 
redshifts between 0 and 0.8 in steps of $z = 0.05$.
Note that we account for the fraction of the initial
stellar  mass that is returned to the interstellar medium by 
evolved stars (e.g., we output the mass of living stars and remnants, not the mass formed);
\item the star formation history, including fraction of stars formed in recent
      bursts, time of truncation etc.;  
\item the metallicity;
\item the dust parameters $\tau_V$ and $\mu$;
\item the stellar velocity dispersion;
\end{enumerate}

The choice of priors is important in Bayesian analysis; we 
test the dependence of our stellar mass estimations 
on the input parameters of the model library in \S5.2.

\subsection{Identifying the significant principal components of the spectral library}
Our method makes use of Principle Component
Analysis (PCA) , a standard multivariate analysis
technique (see Budavari et~al 2009., for a recent discussion). 
A spectrum containing  M wavelength points  
can be regarded as a  single point in an M-dimensional space. 
A group  of spectra form 
a cloud of points in this space. 
PCA searches for a vector (principal component) which has as 
high a variance as possible in the cloud of points.
Each succeeding component in turn has the 
highest variance possible under the constraint 
that it be orthogonal to (i.e. uncorrelated with) the 
preceding components. 

Before we run the PCA code on the library of models, we mask 
the regions around nebular emission lines in the model spectra.
Our models do not include emission lines, and it is important to treat the models
and the real data in exactly the same manner.  
We mask 500 ${\rm km~s}^{-1}$ around the \oii3726.03, 
\oii3728.82,  H$_8$3889.05, \neiii3869.06, 
H$\gamma$4101.73, H$\delta$4340.46, H$\beta$4861.33, 
\oiii4959.91, and \oiii5007.84\AA\ lines. Each spectrum 
in the masked library is normalized to its mean flux between 
$3700-5500$\AA.

\begin{figure}
\bc
\hspace{-0.6cm}
\resizebox{8.5cm}{!}{\includegraphics{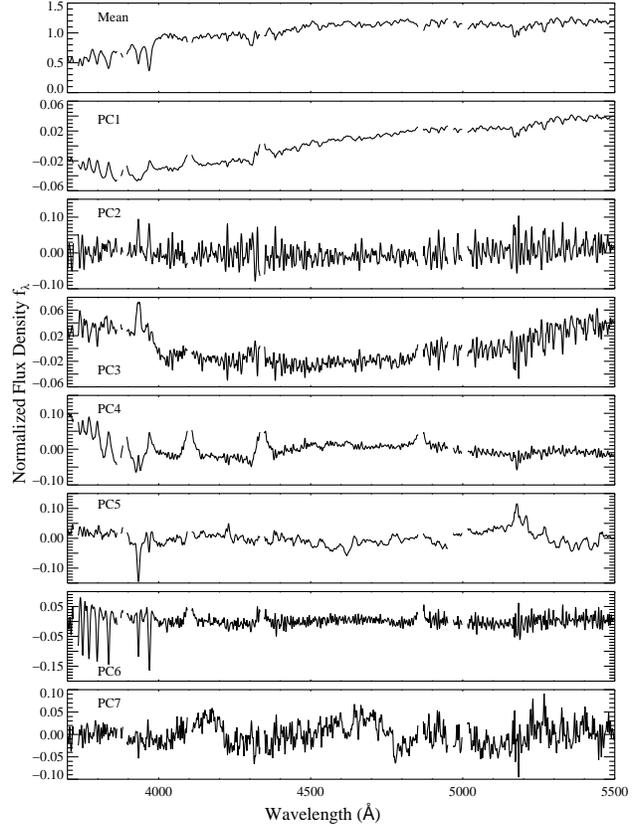}}\\%
\caption{From top to bottom: the mean spectrum of the model library
followed by  the first 
to the seventh eigenspectra.
\label{espec}}
\ec
\end{figure}

Let us  express the $i$-th normalized 
spectrum as $S_{i,k_\lambda}$, where $i$ is an integer running
over the 25,000 model galaxies and $k_\lambda$ is an integer running over each   
pixel in the spectrum. 
We calculate the mean spectrum of the masked library and subtract this from
each of the model spectra. We then run our PCA code on the ``residual" spectra.   
Figure \ref{espec} presents the mean spectrum and the first seven PCs 
for our input model library.

As expected, the mean spectrum is typical of that of a galaxy
with an intermediate age stellar population. The first PC is relatively
featureless. As we will show, it provides a first-order measure of
the age of the stellar population and  it is strongly correlated with both 
4000\AA\ break and Balmer absorption lines strengths.  
The second PC is quite noisy, but as  we will 
show in \S3.4.1, it encodes information about the stellar
velocity dispersion of the galaxy.  The third PC contains
information about  velocity 
dispersion and metallicity. The fourth PC clearly
recovers information contained
in  the  Balmer absorption lines, even though  the line centers have been
masked. CaII (H+K) and Mgb absorption lines are clearly visible in
the fifth PC; as we will show this component carries much information about
galaxy metallicity.  It is difficult to determine what information is encoded in the sixth and
seventh PCs by simple visual inspection.  In the next section, we present a more
quantitative analysis that demonstrates that these two PCs encode  information about 
velocity dispersion and metallicity, respectively.

\subsection{Decomposing each model and observed spectrum into its PC representation}
\subsubsection{Projection of the model library}
The $i$-th model spectrum $S_{i,k_\lambda}$ is projected onto the eigenspectra as
follows: 
\begin{equation}
S_{i,k_\lambda} = M_{k_\lambda} +  \sum_{\alpha} C^m_{i,\alpha} ~E_{\alpha,k_\lambda} + R_{i,k_\lambda},
\end{equation}
where $M_{k_\lambda}$ is the mean spectrum of the model library.   
The integer $k_{\lambda}$ indexes the rest-frame wavelength bin of the
spectrum.
$C^m_{i,\alpha}$ is the amplitude of the $\alpha$-th PC 
$E_{\alpha,k_\lambda}$ (Note that $\alpha$ ranges from 1 to 7, and
the superscript $m$ refers to to the fact that the {\em models} are being projected
onto PC space in this section). 
$C^m_{i,\alpha}$ can be expressed as  
\begin{equation}
C^m_{i,\alpha} = \sum_{k_\lambda}(S_{i,k_\lambda} - M_{k_\lambda}) ~ E_{\alpha,k_\lambda}.
\end{equation}

$R_{i,k_\lambda}$ is 
the residual of the $i$-th spectrum from its PC representation (it
can be regarded as a measure of the theoretical ``noise" in the
decomposition). 
The covariance matrix of the theoretical ``noise" can be written as 
\begin{equation}
Cov^{\rm th}_{k_\lambda, k_\lambda'} =  (1/N_{\rm mod}) \sum_{i=1,N_{\rm mod}}  (R_{i, k_\lambda} R_{i, k'_\lambda})
\end{equation}
where $N_{\rm mod}$ is the number of models in the library (note that this 
noise covariance is indexed 
by rest-frame wavelength). 

Although the PC representation of spectra is compact
and  mathematically convenient,
it is not {\em a priori} clear what physical information is encoded in each of the
components. In some cases (see Figure 2), one can simply eyeball the eigenspectrum
and make an educated guess as to its ``meaning'', but in other cases 
very little can be gleaned from simple visual inspection.            

\begin{figure}
\bc
\hspace{-0.6cm}
\resizebox{8.5cm}{!}{\includegraphics{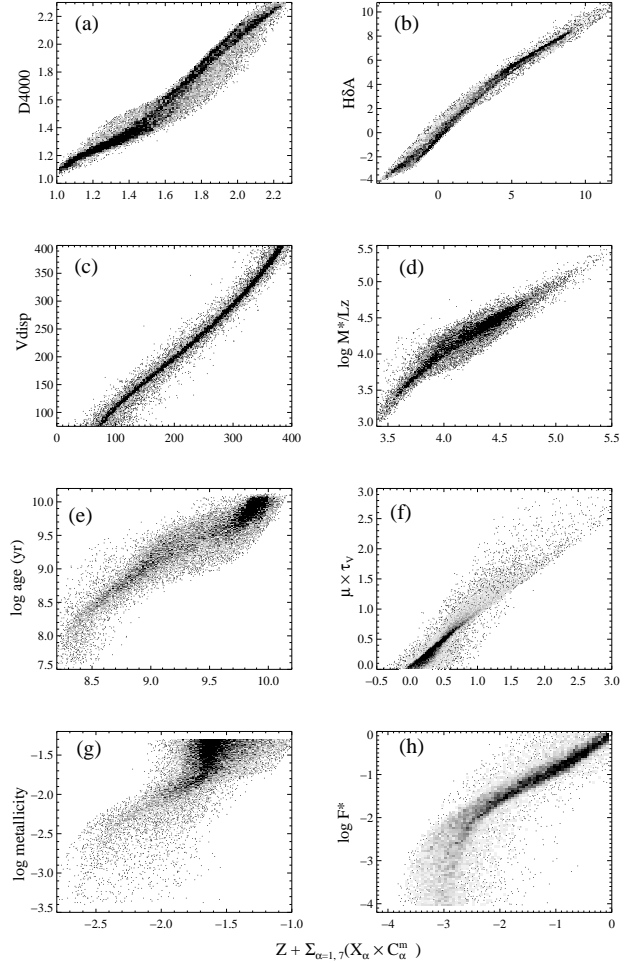}}\\%
\caption{The correlation between 8 different  model galaxy spectral properties or parameters:
(a) D4000; (b) H$\delta_A$; (c) velocity dispersion;
(d) $z$-band  stellar mass-to-light ratio; 
(e) luminosity-weighted age; (f) dust extinction; (g) metallicity; (h) the fraction of stars formed in 
the last Gyr, and the linear combination of $C^m_{\alpha}$ that minimizes the scatter
in the correlation (see text).
The values of the coefficient $X_{\alpha}$ and the zero point $Z$
for each case are listed in Table 1.
\label{coef_cx}}
\ec
\end{figure}

\begin{figure}
\bc
\hspace{-0.6cm}
\resizebox{8.5cm}{!}{\includegraphics{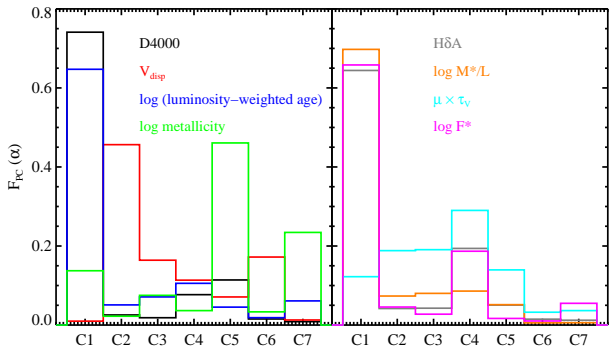}}\\%
\caption{This figure shows the relative contribution of each PC to
the same spectral properties/parameters displayed in Figure 3.
They are color-coded as follows: 
Black $-$ D4000; red $-$ velocity dispersion; blue $-$ luminosity-weighted age; 
green $-$ metallicity; grey $-$ H$\delta_A$; orange $-$ stellar mass-to-light ratio;
cyan $-$ dust extinction; magenta $-$ fraction of stars formed in the last Gyr.
\label{fpc}}
\ec
\end{figure}

In order to develop a better understanding of the information encoded
in the PCs , we search for the best correlation between
$Z + \sum_{\alpha} (X_{\alpha} \times C^m_{\alpha}$)  and 
a variety of galaxy parameters ($P$) that we store for
each  model spectrum   (examples of $P$ include spectral properties such as 
D4000, H$\delta_A$, and velocity dispersion ($V_{\rm disp}$), as well
as model parameters such as 
stellar mass-to-light ratio, light-weighted age, dust 
extinction, metallicity, and  fraction of stars formed in 
the last 1 Gyr ($F_*$)). This can be thought of as finding
the linear combination of the PC amplitudes that
best predicts the  parameter $P$ when averaged over all
the model spectra in the library. The  zero point $Z$ and coefficients 
$X_{\alpha} $ are calculated by minimizing 
\begin{equation}
\Delta = \sum_{i=1}^{N_{\rm mod}} 
\big[\sum_{\alpha} X_{\alpha} \times C^m_{i,\alpha} 
 + Z - P_i \big]^2.
\end{equation}
When we perform this exercise, we  increase the number of PCs used in the  
projection one at a time, each time checking the mean correlation  
between $Z + \sum_{\alpha} (X_{\alpha} \times C^m_{\alpha}$) and $P$
as well as its scatter . We find that our results converge when   
$\alpha = $7; and therefore use  
the first seven PCs in our analysis. 

\begin{table*}
\centering 
\footnotesize
\caption{Values of $X_\alpha$ and $Z$ for each
spectral property or physical  parameter $P$}
\begin{tabular}{lrrrrrrrr}\hline\hline
$P$ & $C_1$ & $C_2$ & $C_3$ &$C_4$ &$C_5$  & $C_6$ & $C_7$ &$Z$ \\ 
(1) & (2) & (3) & (4) & (5) & (6) & (7) & (8) & (9)\\ \hline
D4000                                & 0.049       & $-$0.018 & 0.016         & 0.083        &  $-$0.168  & 0.039            & 0.029        & 1.619     \\
H$\delta_A$                      & $-$0.517 & 0.369       &$-$0.460    & $-$2.551  & 0.905         & $-$0.454      & $-$0.497  & 2.808     \\
$V_{\rm disp}$                  &  0.267      & 135.517  &$-$59.563  & 50.294      &$-$42.942  & $-$176.941 & 17.609      &226.919 \\
log $M_*/L_z$                       &$-$0.051  &$-$0.059  &$-$0.079    & 0.103        &$-$0.084    &0.015             &$-$0.022    &$-$4.209\\
log age(yr)                        &0.071         &$-$0.061  &$-$0.105    & 0.189        &0.111          & 0.078            &0.340          & 9.385     \\
$\mu \times \tau_V$         &0.017         &0.295        &0.366          &$-$0.678   &0.445          &$-$0.176       &$-$0.265    & 0.528     \\
log metallicity                    &0.009         &$-$0.017  &0.070          &$-$0.041   &$-$0.710     &$-$0.088       &$-$0.821   &$-$1.692\\
log $F_*$                           &$-$0.437   &0.333        &$-$0.245    &$-$2.032   &$-$0.248    &$-$0.259        &$-$1.855    &$-$3.631\\
\hline
\end{tabular}
\end{table*}

Figure \ref{coef_cx} shows the correlations  between
$Z + \sum_{\alpha} (X_{\alpha} \times C^m_{\alpha}$)  and
a variety of different galaxy parameters $P$.
We have included D4000 and H$\delta_A$ in this set even though they are not
physical parameters, because they were used
extensively in our previous work and we would like to understand how
they relate to our new system of PCs. In addition, we include   
stellar velocity dispersion, $i$-band mass-to-light ratio, 
$r$-band light-weighted mean stellar age,
metallicity, dust extinction and the fraction of stars formed in the last
Gyr. 
We see that we are able to recover very accurate estimates of
D4000, H$\delta_A$, velocity dispersion  and  dust content 
from the principal components.      
This is not surprising, because the 4000 \AA\ break and the Balmer absorption
lines are the strongest features present in the spectra for the wavelength range
that we have chosen.  Likewise, increasing extinction and velocity dispersion
modify the shape of the spectrum and the width of the spectral features
in a roughly linear way, so it is expected that the correlation with the appropriate
combination of PC components will be very tight. 

On the other hand, there are well
known degeneracies between stellar age and metallicity that affect
many stellar features \citep{oconnel86}. In past work, certain specific features have been
identified as being key to breaking this degeneracy \citep[e.g.,][]{worthey94, vazdekis99, 
mt00, borgne04}, so it is interesting to see
whether our PCA technique is capable of doing the same. In Figure 3, Panels (d)
and (e) indicate that
one is able to recover reasonably  accurate estimates of stellar mass-to-light
ratio and mean stellar age. 
Panel (h) shows that our  method is able to cleanly identify  galaxies
in which more than  $\sim$ 1\% of the stellar population formed in the last Gyr.
Panel (g) shows that metallicity can be recovered for values that are below
solar ($\rm {log~metallicity} < -1.7$). We note, however, that
we have not yet considered the effect of varying element abundance ratios,
which significantly complicates metallicity estimation in real
elliptical galaxies \citep{tmb02}.

In Table 1, we list the set of $X_{\alpha}$ and $Z$ values for each parameter $P$.  
In Figure \ref{fpc}, we attempt to illustrate the relative ``importance''  of
each of the PC components in estimating different galaxy parameters. 
We define $F_{\rm pc}(\alpha)$  as 
\begin{equation}
F_{\rm pc}(\alpha) = \frac{\sum_{i=1}^{N_{\rm mod}} |X_{\alpha} \times C^m_{i,\alpha}|}{ \sum_{\alpha=1}^7 \sum_{i=1}^{N_{\rm mod}} |X_{\alpha} \times C^m_{i,\alpha}|}.
\end{equation}
The results are shown in  Figure \ref{fpc}. For each parameter $P$ in Figure 3,
we plot $F_{\rm pc}$ as a function of $\alpha$, where $\alpha$ is
the index of the PC component. The results are largely
consistent with our previous  discussion of Figure \ref{espec}.   
Information about D4000, H$\delta_A$, stellar mass-to-light ratio, light-weighted age and the fraction of
stars formed in the last Gyr is primarily contained in  PC1,  
with lesser contributions from PC4 and PC5.  
As we have discussed, PC1 provides a measure of the continuum shape,
whereas  PC4 is similar to the component defined in Wild et al (2007)
that provides a measure of the ``bursty'' nature of the past star formation history.
Information about velocity dispersion is encoded  in PC2, PC3 and PC6.
Information about metallicity is mainly contributed by PC5 and PC7.
Interestingly, PC components 2 through 5 contribute almost equally in the
estimation of stellar mass-to-light ratio.

\begin{figure*}
\bc
\includegraphics[angle=0,width=0.9\textwidth]{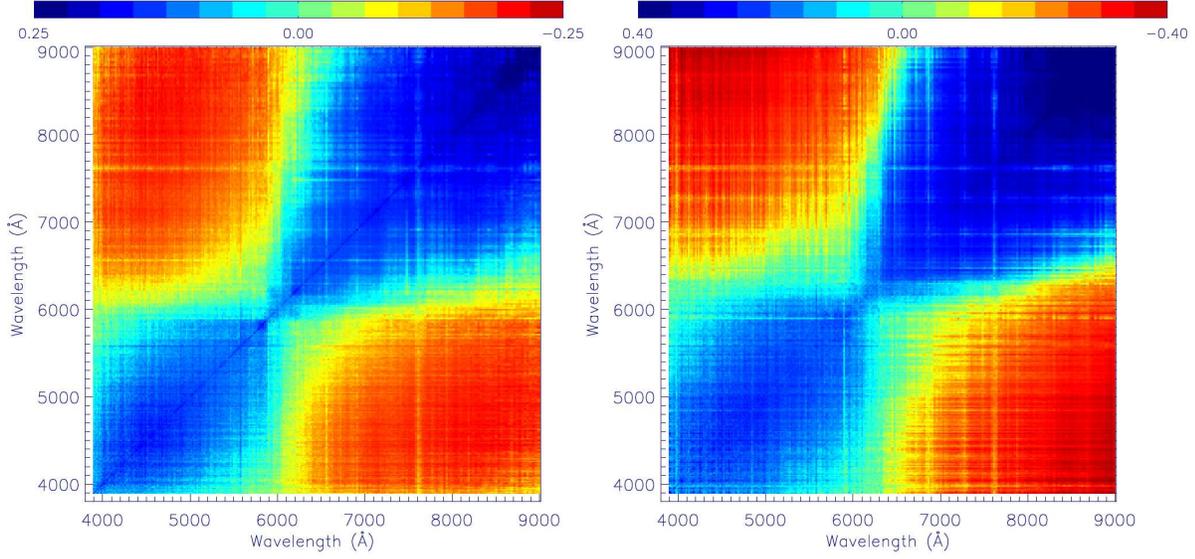}
\caption{The rescaled mean observational error covariance 
matrix $Cov_{l_\lambda,l_\lambda'}$ in the wavelength interval of 
$4000-9000$\AA\ derived from DR7 repeat spectra (left) and BOSS
repeat spectra (right).
\label{cov}}
\ec
\end{figure*}

\begin{figure*}
\begin{center}
\includegraphics[angle=0,width=5.5cm]{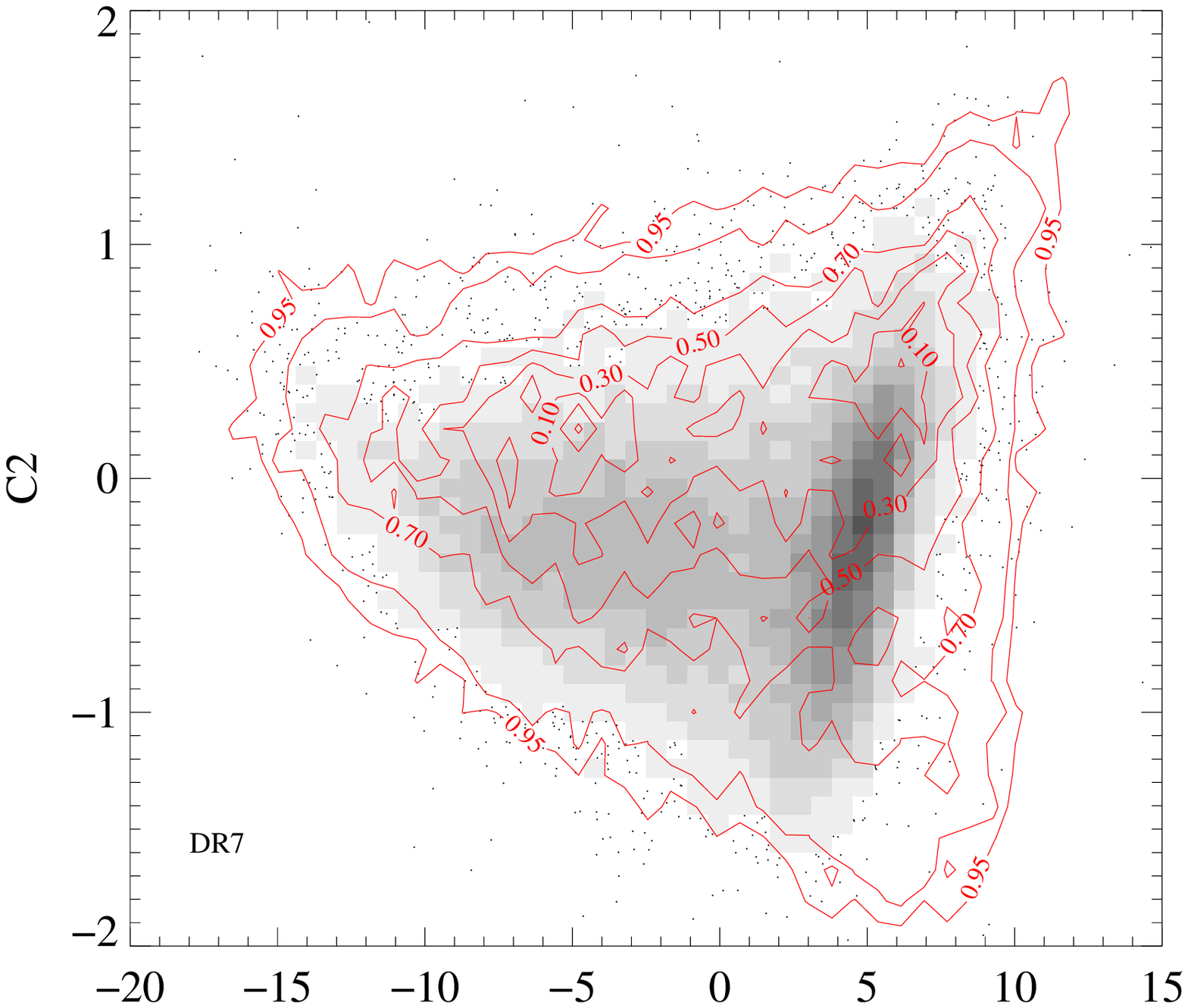}
\includegraphics[angle=0,width=5.5cm]{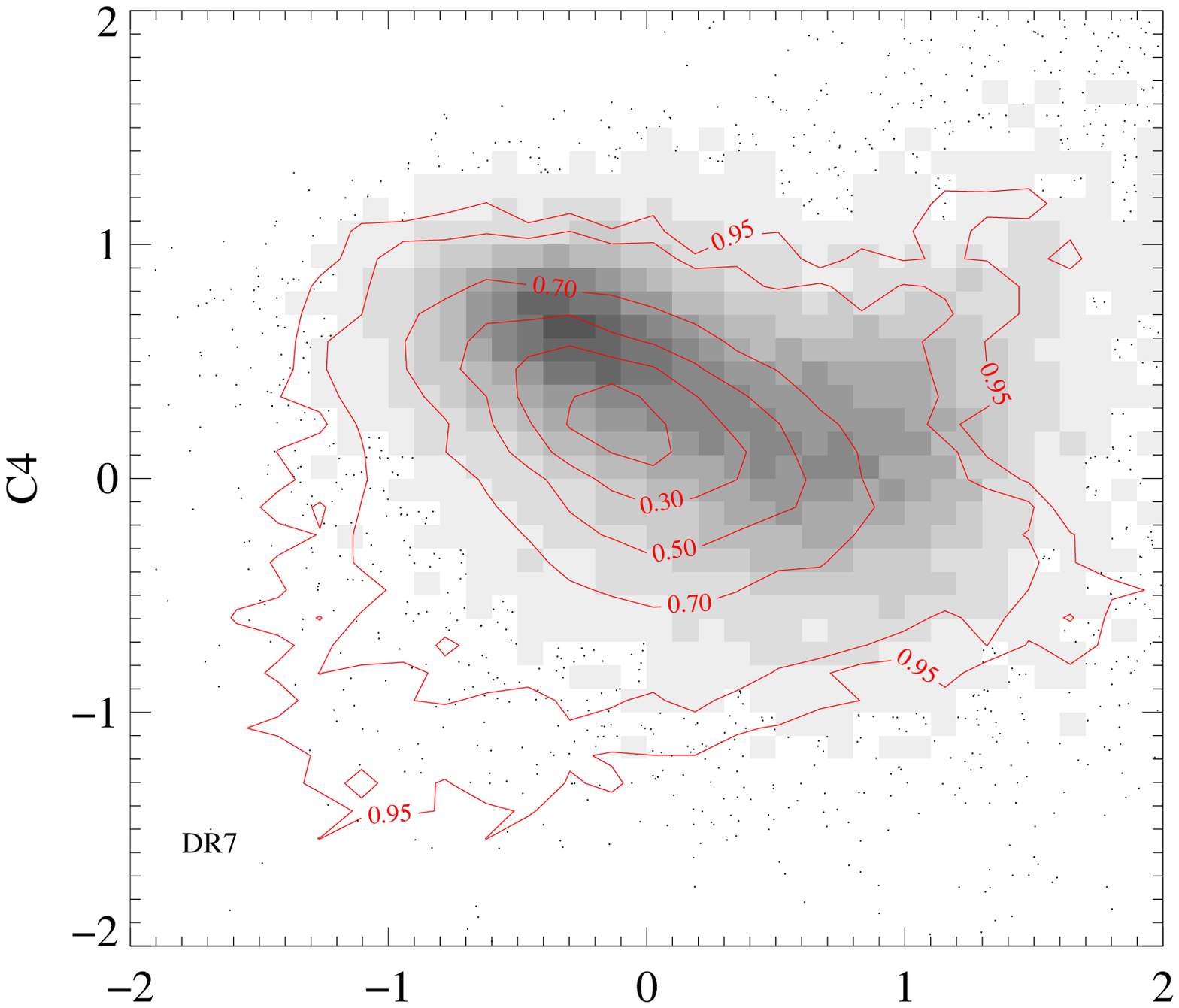}
\includegraphics[angle=0,width=5.5cm]{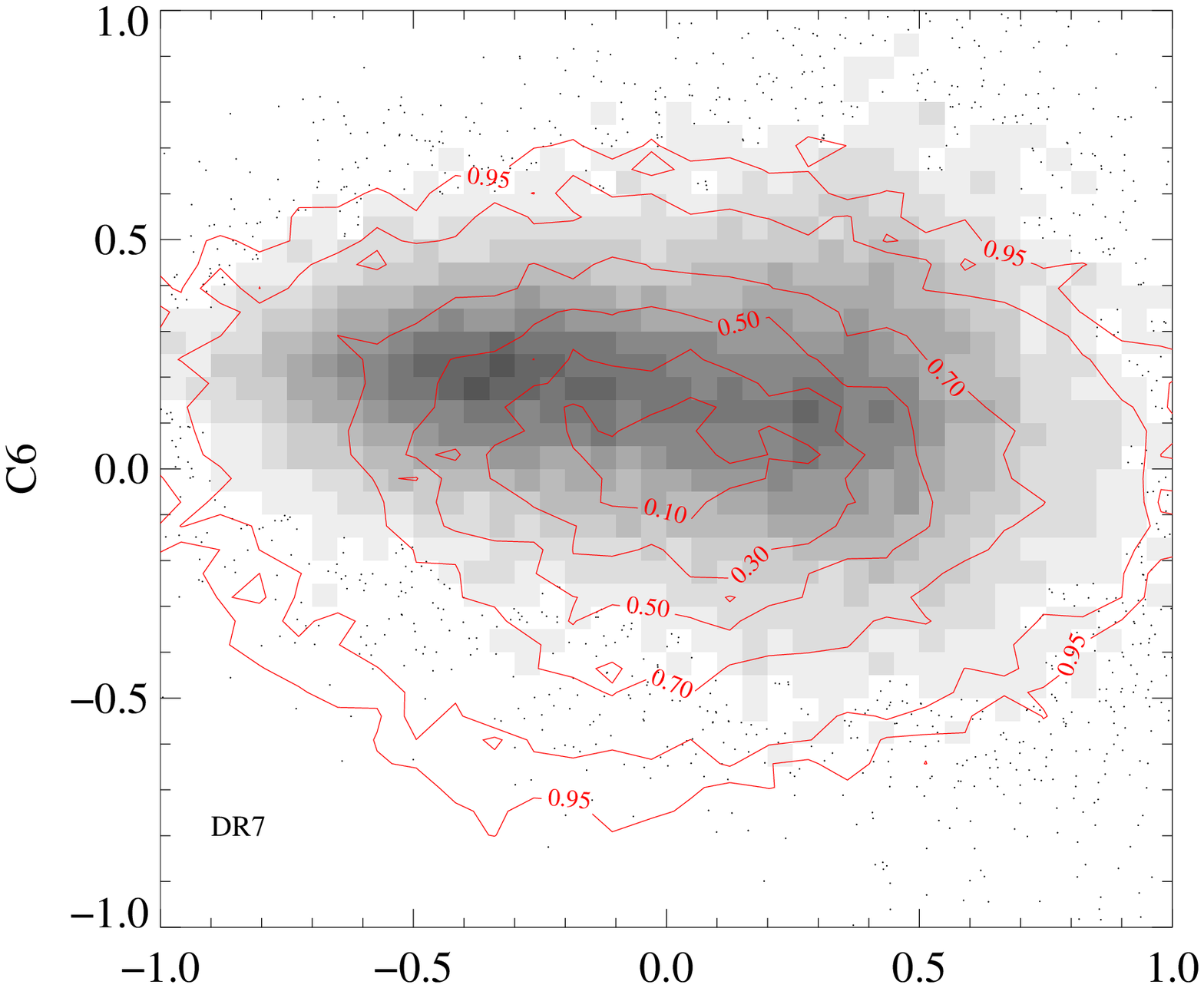}
\includegraphics[angle=0,width=5.5cm]{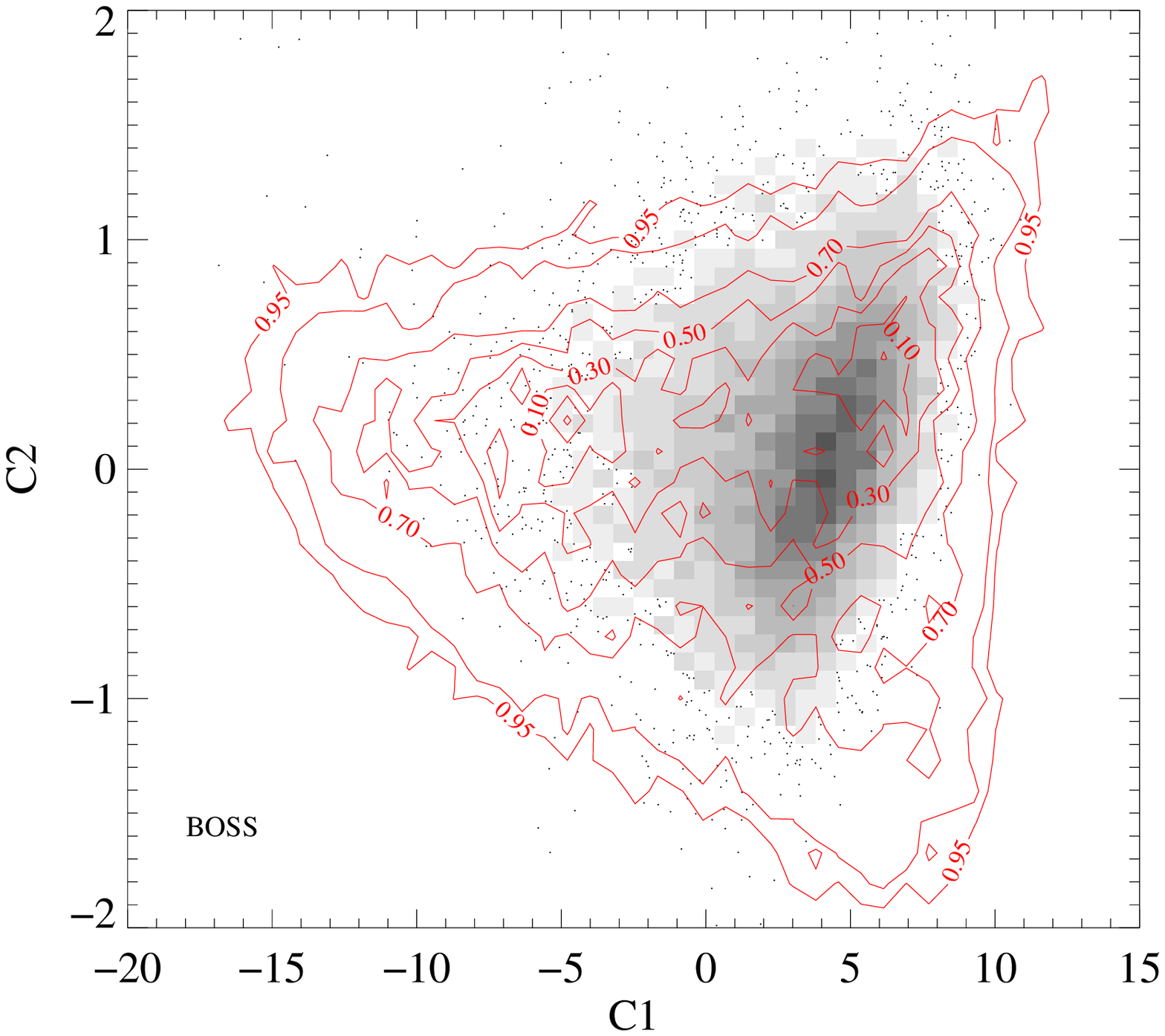}
\includegraphics[angle=0,width=5.5cm]{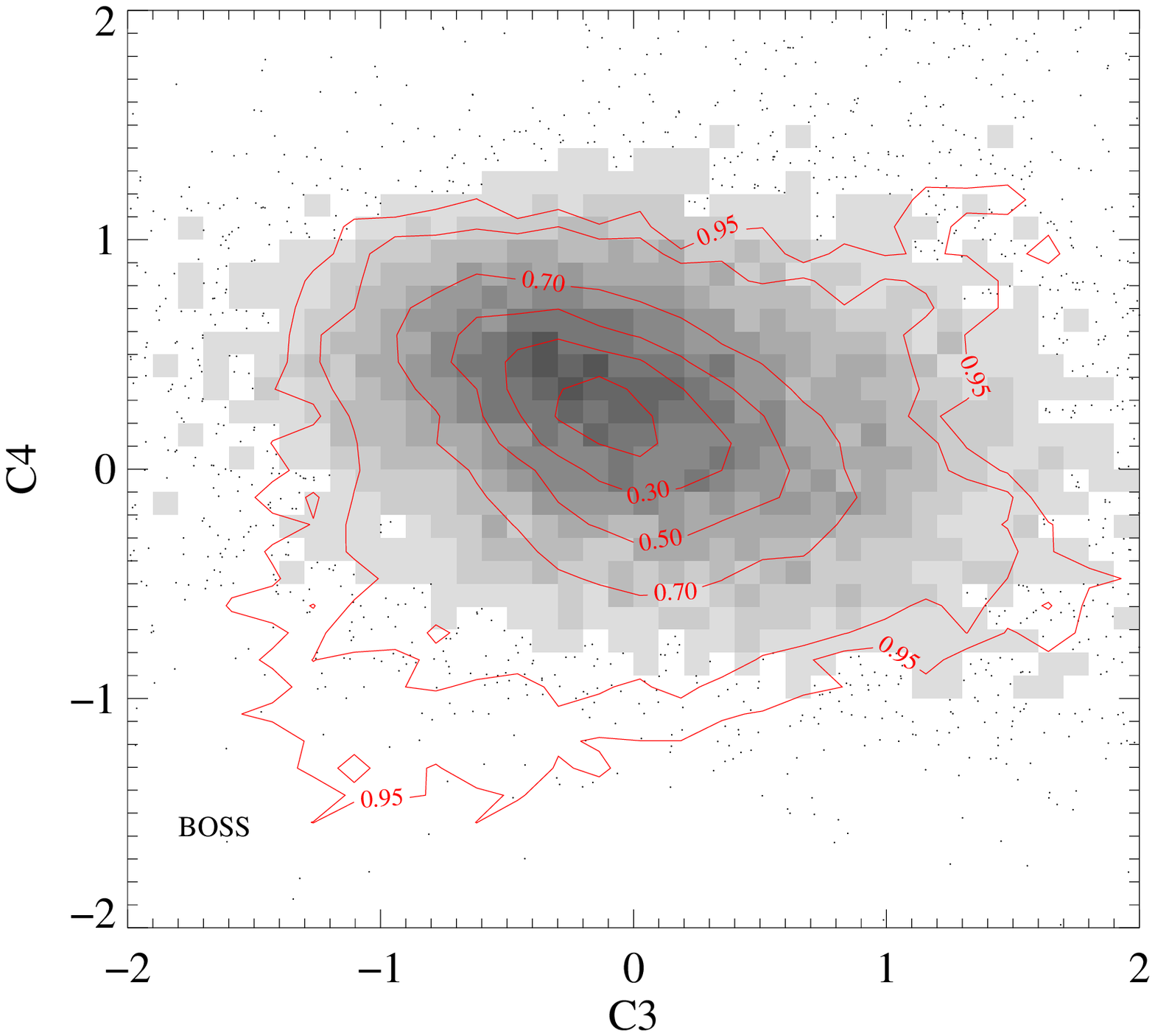}
\includegraphics[angle=0,width=5.5cm]{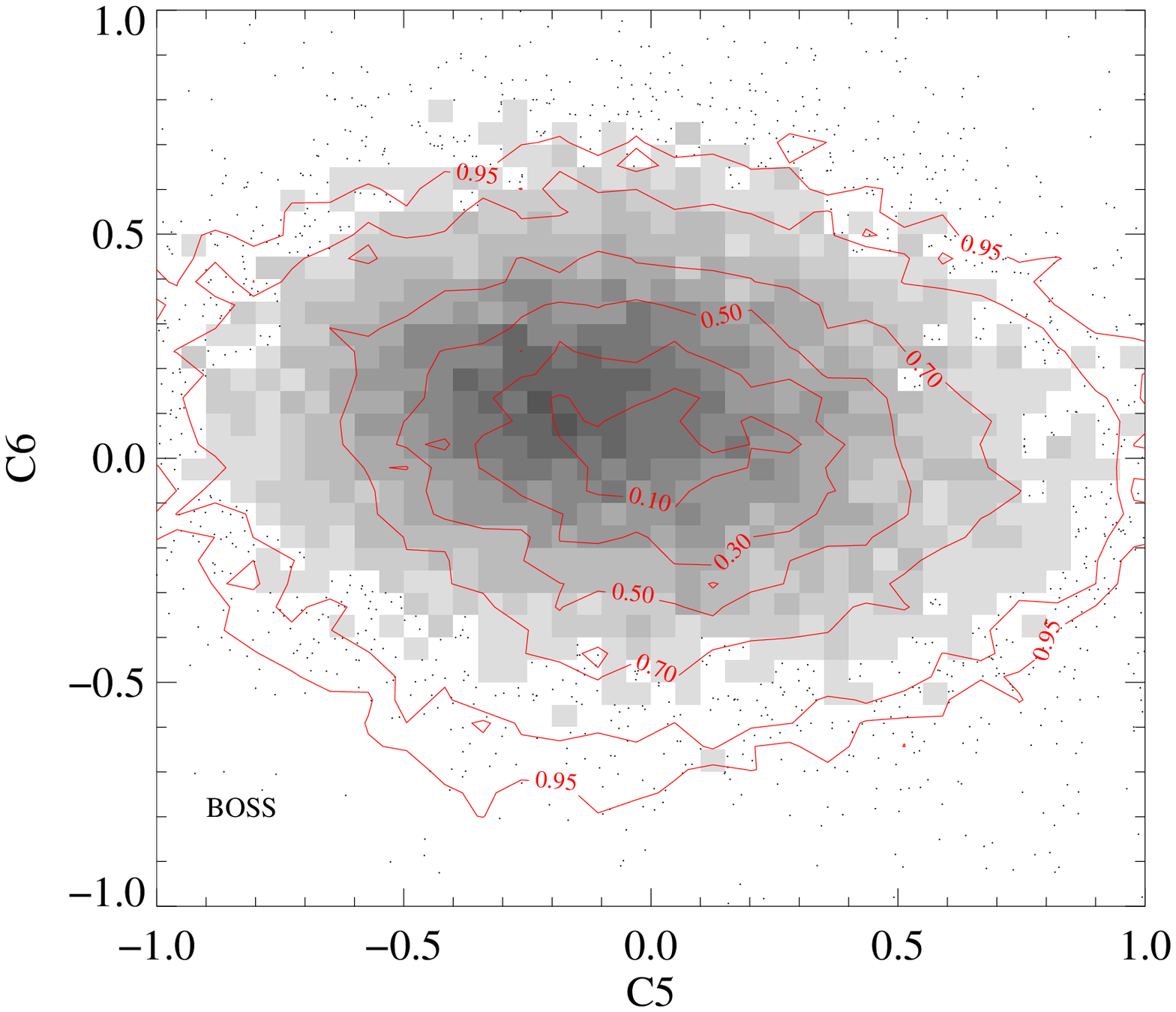}
\end{center}
\caption{This figure shows $C_1$ vs.  $C_2$, $C_3$ vs.  $C_4$ and  $C_5$ vs.  $C_6$ for DR7 galaxies
(top) and for BOSS galaxies (bottom). Models are shown in red contours, while data is shown
in grey scales. We have convolved the model spectra with the errors 
appropriate for DR7 and CMASS when generating the PCA components
for this plot. The outermost contour encloses 95\% of the models.
\label{model_coverage}}
\end{figure*}

\subsubsection{Projection of the real data}
In this section, we describe how we apply the PCA methodology to observed galaxy spectra.
The steps are as follows:
\begin{enumerate}
\item The galaxy spectra are corrected for foreground Galactic 
extinction and the wavelength scale is 
shifted from vacuum to air to match the models.
\item The set of emission lines listed in \S3.3 are masked.
In addition, we  found it necessary to mask $500{\rm km~s}^{-1}$ around  
\hi3770.63, 3797.90, 3835.38, 3889.049, 
3970.072, 4101.734, 4340.464, 
\hei4387.93, 5047.74, \heii5015.68, \neiii3967.79, \oiii4364.21\AA\ lines
in the subset of strong emission line galaxies with EQW of H$\beta$ $< -$5 (12\% of DR7, 1\% of CMASS). We 
have checked that our results are robust to the choice of mask size.
\item  The observed spectrum and its error array are normalized by
dividing by the  flux density averaged over the full observed wavelength range. 
We then apply an integer pixel offset to shift them from observed to rest-frame wavelength\footnote{This is possible because the wavelength 
interval of SDSS spectra is a constant in log-space with $\Delta{\rm log}~\lambda = 10^{-4}$. We rebin the model
spectra to the same wavelength interval.}. 
We denote the normalized flux density and its error arrays as $O_{k_\lambda}$ and 
${\rm Eps}_{k_\lambda}$, respectively.  For ``bad" pixels and night sky lines 
identified in the SDSS mask array, we set the pixel values
in the observed normalized spectrum, $O_{k_\lambda}$, proportional to the value in the mean 
spectrum of the model library, $a \times M_{k_\lambda}$, where $a$ is the mean flux of $O_{k_\lambda}$ between rest frame 
$3700 - 5500$\AA, which takes the different normalizations between models and data into account. (We choose this normalization 
of the data for the convenience of estimating the mean observational error covariance matrices, see Eq. 9). 
The corresponding pixel values in the error array  are set to 10 times the mean error of all the good pixels.  
\item The coefficients of the PC components of the observed spectrum are
\begin{equation}
C_{\alpha}^d = \sum_{k_\lambda} (O_{k_\lambda}
- a \times M_{k_\lambda}) ~E_{\alpha,k_\lambda}.
\end{equation}
Note that the index $k_\lambda$ ranges over the 
pixels between 3700$-$5500\AA\ in the rest frame. 
\end{enumerate}

\subsubsection{Error estimation for the PC coefficients $C_{\alpha}^d$}

This section describes how we estimate the errors on $C_{\alpha}^d$ as 
calculated in equation (7).
 Let us write the observed spectrum over the restricted 
wavelength range we are modeling as
\begin{equation}
O_{k_\lambda} = a \times M_{k_\lambda} + \sum_{\alpha} C^{true}_\alpha~E_{\alpha,k_\lambda}
               + N^{th}_{k_\lambda} \times a + N^{obs}_{k_\lambda}
\end{equation}       
where the first two terms in equation (8) represent the  ``true"  PC representation 
of the galaxy in the absence of errors of any sort.
$N^{th}_{k_\lambda}$ and $N^{obs}_{k_\lambda}$ are independent Gaussian noise vectors
representing the ``theoretical'' and the ``observational'' errors. These two noise vectors
have covariance matrices  
$Cov^{th}_{k_\lambda,k'_\lambda}$ and $Cov^{obs}_{k_\lambda,k'_\lambda}$, respectively. 

$Cov^{th}_{k_\lambda,k'_\lambda}$  
 is the same for each observed galaxy and is given by equation (4). 
The covariance matrix of the  observational errors differs from one galaxy to 
the next. The diagonal terms are given by    
${\rm Eps}^2_{k_\lambda}$, the square of the normalized error array 
of the particular galaxy in question. The off-diagonal terms are somewhat  tricky
to evaluate, because they depend on details of the instrumental response, flux
calibration errors, etc.       

Both the DR7 main galaxy sample and the BOSS CMASS  sample include a large number of 
galaxies that were observed multiple times, due to overlaps between different
spectroscopic fiber plug plates.
These ``repeat spectra'' are extremely useful  for assessing uncertainties in
a variety of estimated quantities. 
We identify a sample of 48,000 ``repeat galaxy spectra'' for  DR7
(19,000 for  CMASS); these two samples are used to construct the  
mean observational error covariance 
matrices for DR7 and CMASS galaxies as 
\begin{equation}
Cov_{l_\lambda,l_\lambda'} =\\ 
\frac{1}{2N_{\rm pair}} \sum_{j=1,N_{\rm pair}}  (O_{j,1,l_\lambda} - O_{j,2,l_\lambda}).
                                       (O_{j,1,l_\lambda'} - O_{j,2,l_\lambda'})
\end{equation}
where $O_{j,a,l_\lambda}$ is the normalized flux of the $a$-th observation ($a$ = 1 or 2) 
of the $j$-th pair. Bad pixels are replaced by the median of  200 neighbouring pixels. 
$l_\lambda$ ranges over all the pixels in the spectrum.
Note that in contrast to the index $k_{\lambda}$ used above, $l_{\lambda}$
indexes wavelength in the observed frame, and 
 $O_{j,a,l_\lambda}$ refers to the observed spectrum. 

In Figure \ref{cov}, we show a diagrammatic representation of the
observational covariance matrices from the DR7 and BOSS repeat observations 
over the wavelength interval of $4000-9000$\AA.  
We have stretched the scale in order to distinguish positive from negative values and
to have a near-logarithmic scale near zero (in practice, we plot
sign($Cov_{l_\lambda,l_\lambda'}$)$|Cov_{l_\lambda,l_\lambda'}|^{0.3}$).  
The absolute values of the elements on the BOSS mean 
observational error covariance matrix are much larger 
than that for  DR7, reflecting the fact that the  
BOSS data are at lower S/N.
The strong shift in sign at 6000 \AA\ reflects the change between the blue and the
red channels of the SDSS spectrograph.
When using the matrices below (see equation 11),  the off-diagonal terms 
of $Cov^{obs}_{k_\lambda,k'_\lambda}$
are set to be the values of  $Cov_{l_\lambda,l_\lambda'}$ at the 
corresponding observer-frame wavelength.

The difference between ``true" and estimated PC coefficients is
\begin{equation}
{\rm d}C_{\alpha} = C^{true}_\alpha - C^d_{\alpha} 
      = \sum_{k_\lambda} E_{\alpha,k_\lambda}~(N^{th}_{k_\lambda} \times a + N^{obs}_{k_\lambda})
\end{equation}  
The covariance matrix of this difference  (i.e the covariance
in the errors on the different  PC coefficients) is then given by 
\begin{eqnarray}
\begin{array}{lcl}
Cov^{\rm pc}_{\alpha,\alpha'} = < {\rm d}C_{\alpha}~ {\rm d}C_{\alpha'}>  = \\
                                                      \sum_{k_\lambda, k'_\lambda} [E_{\alpha,k_\lambda} \times (Cov^{\rm th}_{k_\lambda, k'_\lambda} \times a^2 + 
                                                           Cov^{\rm obs}_{k_\lambda,k'_\lambda}) \times E_{\alpha',k'_\lambda}]
\end{array}
\end{eqnarray}
\begin{figure*}
\bc
\includegraphics[angle=0,width=0.48\textwidth]{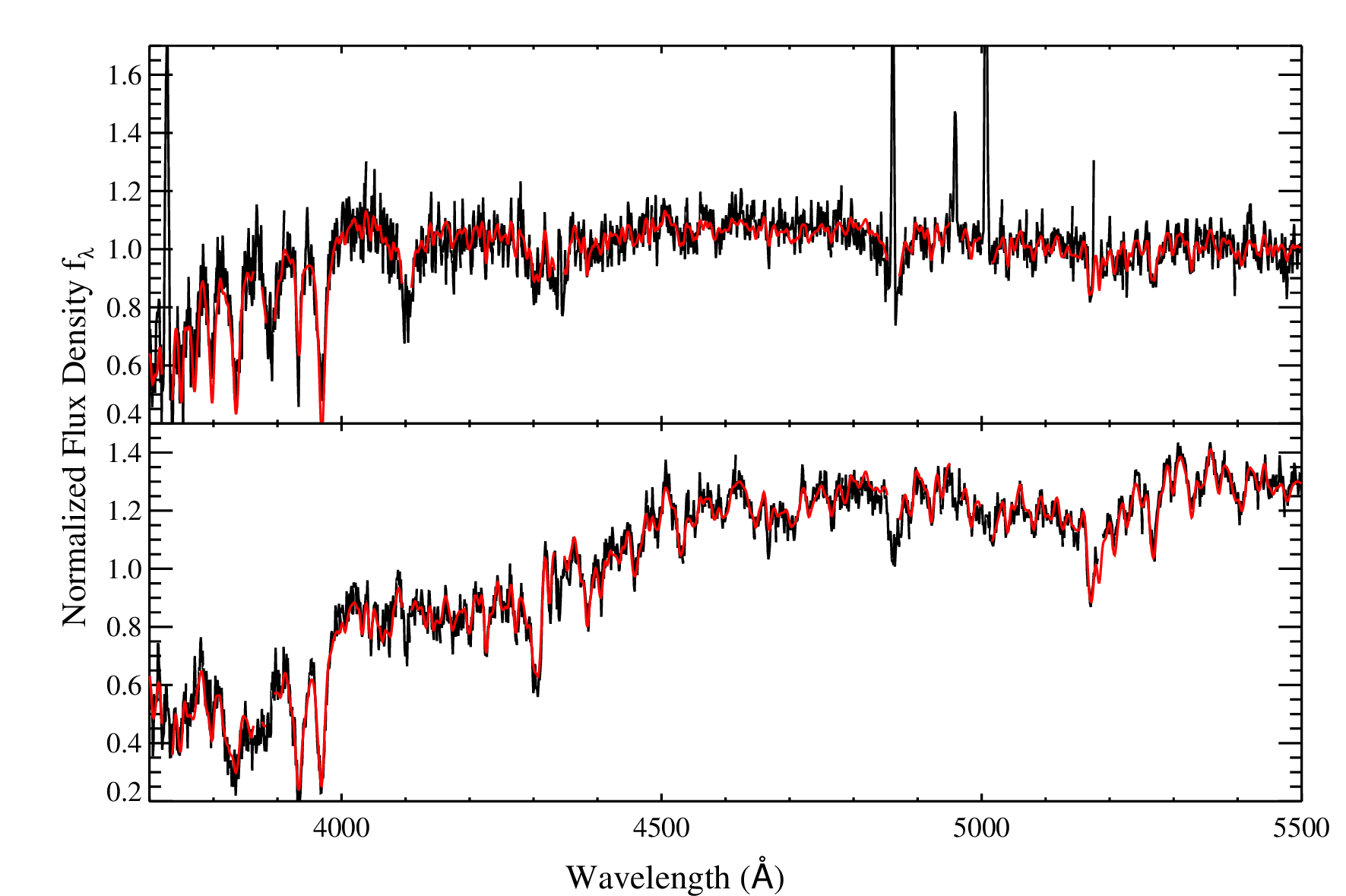}
\includegraphics[angle=0,width=0.48\textwidth]{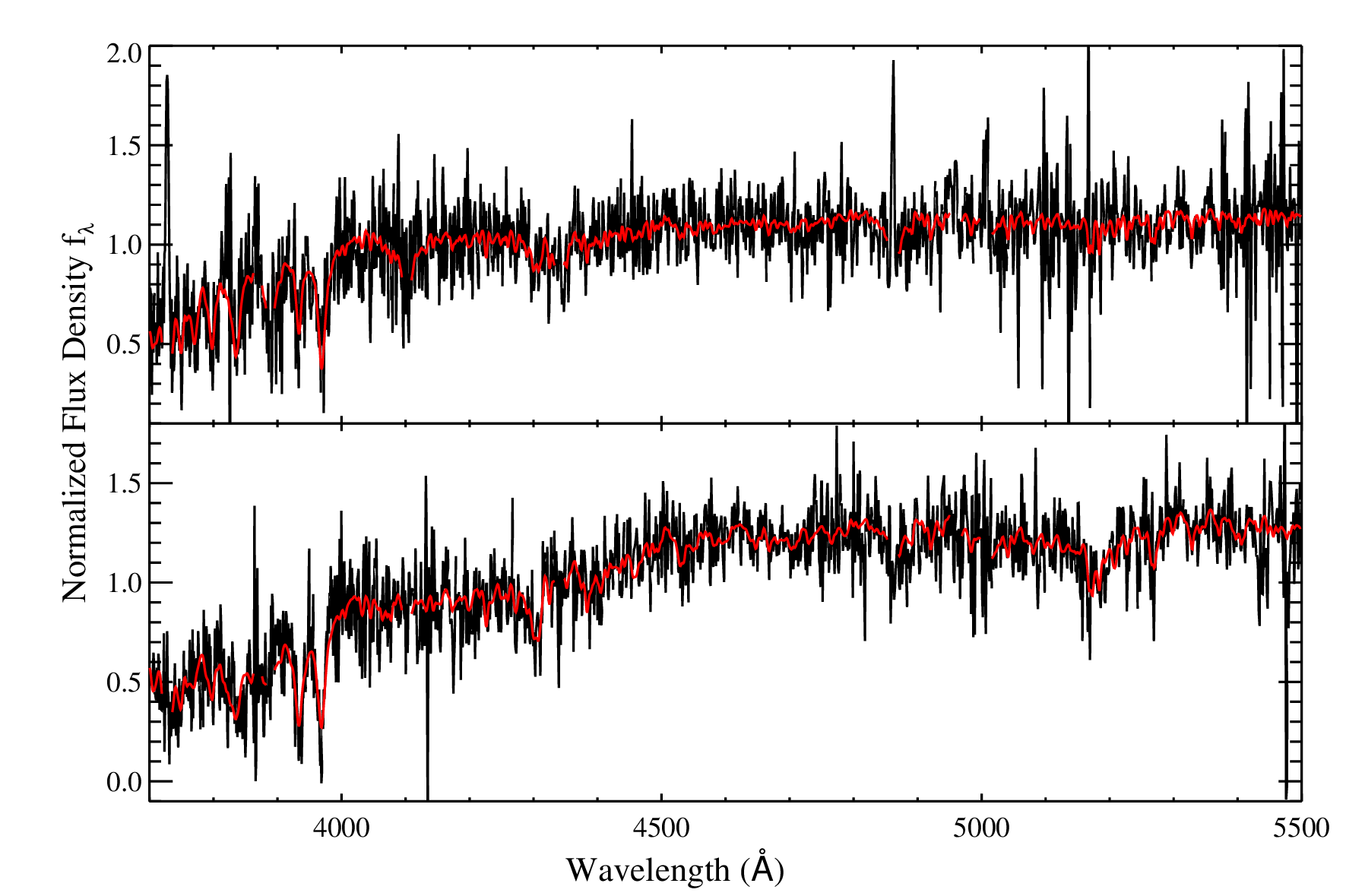}
\caption{Left panel: Example DR7 spectra (black) and the corresponding PCA fits (red) 
for a late-type galaxy (top) and 
an early-type galaxy (bottom). The spectrum is plotted
over the wavelength interval $3700 - 5500$\AA. Right panel: 
Two example spectra and fits from BOSS. Note emission lines are not fit.}
\label{fit}
\ec
\end{figure*}

\subsection{ Comparison of the PC components derived for the models and for the data}
If parameter estimation using the  model library is to be robust, we must  make sure
that the models  cover the same range of PC space as the real data.
Figure \ref{model_coverage} shows 
$C_1$ vs.  $C_2$,  $C_3$ vs.  $C_4$ and $C_5$ vs. $C_6$ for DR7 galaxies
(top) and for BOSS galaxies (bottom). Models are shown in red contours, while data are shown
in gray scales. We have convolved the model spectra with the errors 
appropriate for DR7 and CMASS when generating the PCA components
for this plot, so that the comparison between
models and observations is realistic.   As can be seen, 
data and models cover roughly the same regions of
PC parameter space for both the DR7 and the CMASS samples.    
We note that the worst discrepancies between models and data are for the
$C_4$ index, which encodes information about Balmer absorption lines. 
This problem was previously noted by \citet{wild07} in their analysis
of post-starburst galaxies using data from the SDSS Data Release 4 \citep[DR4;][]{abazajian05}.
Interestingly, agreement between models and data in the $C_4$ versus $C_3$
place is significantly better for the BOSS galaxies, which are significantly
more massive and have metallicities close to solar, where the coverage
by the stellar libraries is more complete.   

\subsection{ Estimation of physical parameters and their uncertainties}

For an observed galaxy at redshift $z$, we select only models that have an age smaller than 
the age of the universe at that redshift. We step through the models one at a time, calculating 
the $\chi^2$ as follows:                                                                
\begin{equation}
\chi_i^2 = \sum_{\alpha, \alpha'} (C^{m}_{i,\alpha} - C^d_{\alpha}) P_{\alpha,\alpha'} (C^{m}_{i,\alpha'} - C^d_{\alpha'}) 
\end{equation}
where $P_{\alpha,\alpha'} $ is the inverse  of $Cov^{\rm pc}_{\alpha,\alpha'}$. 

We define a weight  $w_i = {\rm exp}(-\chi_i^2/2)$ to describe the similarity between the 
given galaxy and model $i$. We then build a   
probability distribution function (PDF) for
each parameter $P$, by looping over all the model galaxies in the library and   
by summing the weights $w_i$  at the value of $P$ for each model.            
We normalize the final PDF and note the parameter  values at the 
2.5, 16, 50 (median), 84 and 97.5 percentiles of the cumulative PDF. We adopt the 
median as our nominal estimation of $P$ and the $16 - 84$ percentile range of the PDF as 
its  $\pm1 \sigma$ confidence interval.

\begin{figure}
\bc
\hspace{-0.6cm}
\resizebox{8.5cm}{!}{\includegraphics{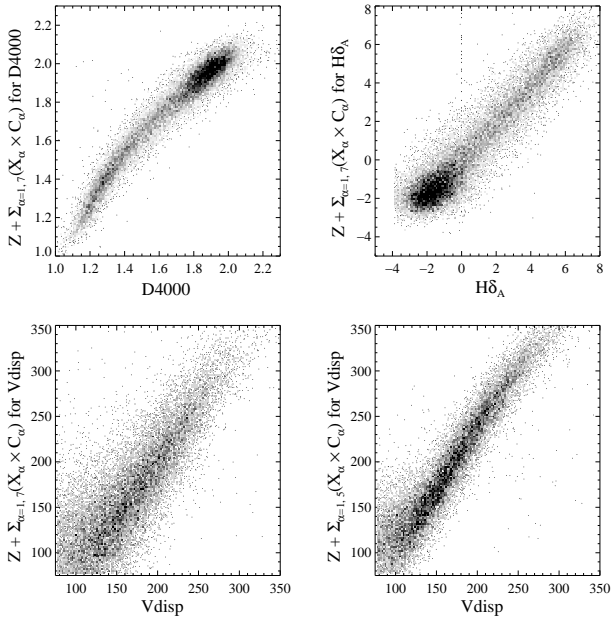}}\\%
\caption{ The linear combinations of the PC coefficients predicted to provide
the best representation of D4000, H$\delta_A$ and stellar velocity dispersion
(see Section 3 for details)
are plotted as a function of direct measurements of these quantities from the SDSS pipeline.  
5 PCs are used for the bottom-right panel, while 7 PCs are used for the other three panels.
The sample has been restricted to DR7 galaxies with spectral
 median S/N per pixel greater than 10. 
\label{pc_vs_lick}}
\ec
\end{figure}

\section{Advantages of the Principal Component Method} 

In the previous section, we described our methodology for deriving a set of
principal components from a library of model spectra, for decomposing real
galaxy spectra into linear combinations of these components, and for
estimating  errors on the derived PC  amplitudes. We also outlined a Bayesian
technique for parameter estimation using the input model library.
In this section, we will attempt to illustrate the power of our methodology
by means of  some scientific applications.

\subsection {A robust method for stellar continuum fitting} 

To decompose the  galaxy spectrum into the emission produced by 
ionized gas and that produced by stars, one usually attempts to fit a model 
that describes the stellar component of the galaxy and then one ``subtracts''
this model from the observed spectrum. Certain emission lines, for example
H$\beta$, frequently occur in deep absorption troughs, particularly in early-type
galaxies, so it is important  that the model for the stellar continuum
be as accurate as possible. 

In previous work \citep{brinchmann04, tremonti04}, we employed a
template-fitting procedure to model the stellar continuum. As described
in section 3, instead of a set of
templates, our new method  finds the linear combination
of PC eigenspectra that best fits the stellar continuum.      
In Figure \ref{fit}, we show two examples of PCA fits to  DR7 (left) 
and BOSS (right) galaxy spectra over the wavelength interval from 3700 to 5500\AA. 
The upper panel shows the spectrum of a star forming galaxy 
with strong Balmer emission and small 4000 \AA\ break, 
while the lower panel shows a typical 
early-type galaxy with strong stellar absorption features. The black lines are the normalized 
observations, the red lines show the best PCA fits, which are clearly very good.

\subsection {Improvement in S/N over Lick Indices when using PC-based techniques}  

\begin{figure*}
\bc
\hspace{-1.6cm}
\resizebox{13cm}{!}{\includegraphics{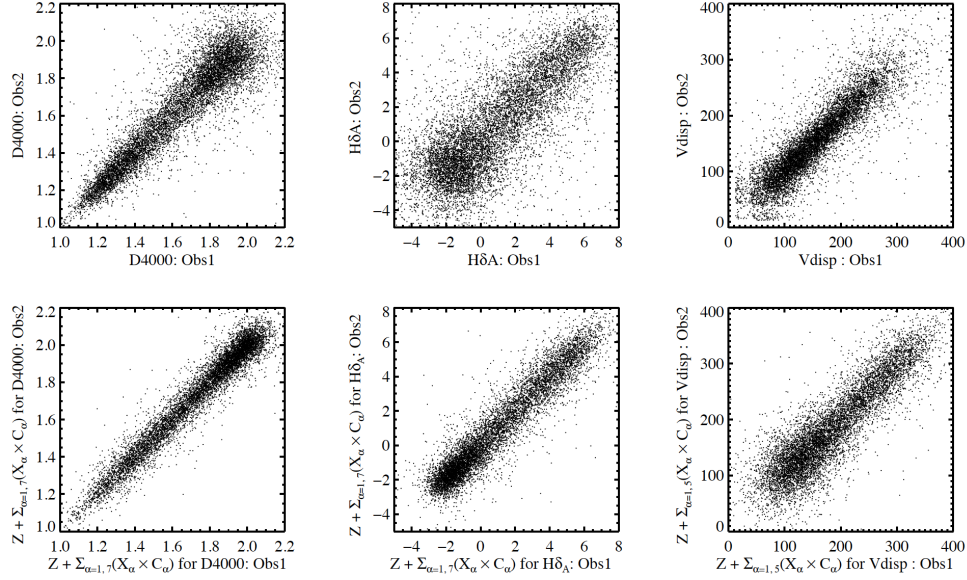}}\\%
\caption{A sample of DR7 galaxies with two observations has been extracted
from the SDSS database and   measurements of D4000 (top-left) , 
 H$\delta_A$ (top-middle) and   $V_{\rm disp}$
(top-right) are plotted against each other in the top
panel. The corresponding PC representations
of these quantities are plotted against each other in the bottom panel.  
\label{repeat_dr7}}
\ec
\end{figure*}

\begin{figure*}
\bc
\hspace{-1.6cm}
\resizebox{13cm}{!}{\includegraphics{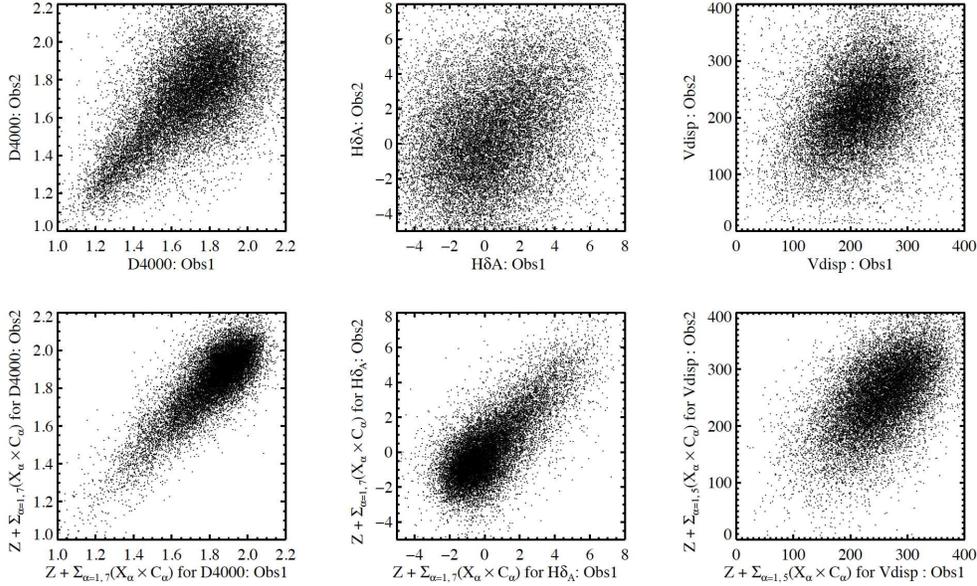}}\\%
\caption{This figure is similar to Figure 9, except that results are shown for multiply-observed CMASS galaxies.  
\label{repeat_boss}}
\ec
\end{figure*}

It is traditional to focus on specific stellar absorption features 
over a narrow wavelength interval when analyzing the
ages and metallicities of the stellar populations of 
galaxies from their spectra. For galaxies with old stellar populations, 
the Lick/IDS system of $\sim$25 narrow-band indices 
is often used \citep{worthey94, wortheyott97, gorgas99}. For actively
star-forming galaxies, the 4000 \AA\ break \citep{balogh99} and Balmer absorption
line features, such as the H$\delta_A$ index, provide important information
about stellar age and recent star formation history \citep{kauffmann03a}.  
Finally, the velocity dispersion of the stars in the galaxy  
is traditionally estimated by comparing the width of stellar absorption 
features with those in unbroadened stellar templates (see Appendix B of Bernardi et al. 2003
for a review).

In Figure 3 of Section 3, we used {\em model galaxy spectra} to  illustrate how narrow-band
spectral indices such as D4000 and H$\delta_A$ could be ``recovered'' from appropriate
linear combinations of the PCs. We also showed that the PCs could in principle
be used to recover an estimate of stellar velocity dispersion.   
Figure \ref{pc_vs_lick} illustrates how well this works for real
galaxy spectra.   We compare $Z + \sum_{\alpha} (X_{\alpha} \times C_{\alpha})$
(using the $X_{\alpha}$ and $Z$ values in Table 1)  with actual measurements of 
D4000, H$\delta_A$ and stellar velocity dispersion for a subset of galaxies
drawn from the DR7 sample (note that these measurements are drawn from
the MPA/JHU database). Since the purpose of this figure is to illustrate  how well
the recovery of traditional Lick indices and stellar velocity dispersion estimates
is able to work {\em in principle}, we only plot galaxies with spectra
where the  median S/N per pixel is greater than 10. 

We obtain very tight correlations between the appropriate linear combinations 
 $Z + \sum_{\alpha} (X_{\alpha} \times C_{\alpha})$   
and both D4000 and H$\delta_A$. We were not able to 
recover a good correlation with the velocity
dispersion from the SDSS pipeline unless we reduced the 
number of PC components from 7 to 5.
As illustrated in Figure \ref{fpc}, PC$_6$ should in principle
contribute significantly to our estimate of  $V_{\rm disp}$. 
The increase in  scatter when using seven PCs instead of five 
is attributable to the fact that  
there are large uncertainties in measuring PC$_6$ even for 
spectra with S/N per pixel greater than 10. Below, we will always use just 5 PCs
when estimating velocity dispersions, but 7 PCs
when estimating all other quantities.

We now demonstrate that  $Z + \sum_{\alpha} (X_{\alpha} \times C_{\alpha})$ has higher
signal-to-noise ratio than the direct measurements of D4000, H$\delta_A$ and velocity
dispersion for {\em typical} galaxy spectra drawn from the DR7 and CMASS samples.
As we will show, the improvement is most striking for the low S/N CMASS spectra.
The upper panel of Figure \ref{repeat_dr7} illustrates the scatter obtained between 
the directly measured values of  D4000, H$\delta_A$, and velocity dispersion
for two different  observations of the same galaxy in DR7. The bottom panel
shows the same format for the corresponding  $Z + \sum_{\alpha} (X_{\alpha} \times C_{\alpha})$
representation of the same quantities. 
Figure \ref{repeat_boss} displays the same comparison based on repeat observations of  
CMASS galaxies; note that for this sample,
$> 85\%$ galaxies have a median S/N per pixel below 4.
The improvement in S/N exhibited by the PC representations of D4000 and H$\delta_A$ 
are highly significant both for the DR7 galaxies and the CMASS galaxies.
The improvement in the PC-based velocity dispersion estimate is 
only obvious for the low S/N CMASS sample. 

\section{Application of the Principal Component Method to Stellar Mass Estimation} 

The application we wish to highlight in this paper (more applications will follow
in subsequent work) is the derivation of robust PCA-based stellar masses for galaxies
from the DR7 and CMASS samples. 
For each sample, two sets of stellar masses are estimated: the stellar mass measured within
the fiber and the total mass. In the case of
DR7, the masses are calculated by multiplying $M_*/L_z$ by 
the $z$-band luminosity measured within the $3^{\prime\prime}$ 
SDSS-I spectrograph 
fiber aperture, or by the luminosity $L_z$ derived from the 
SDSS $z$-band ``model'' magnitude. 
For the CMASS sample, $M_*/L_{i}$ is multiplied by the
$i$-band luminosity measured within a $2^{\prime\prime}$ diameter aperture
matched to the smaller BOSS spectrograph fibers, or
by the  $i$-band luminosity $L_i$ derived from the SDSS $i$-band ``cmodel'' magnitudes. 
We note that ``model'' and ``cmodel'' magnitudes are close to equivalent
for DR7 galaxies, but diverge at the fainter magnitudes of the CMASS sample.  In general ``model'' magnitudes
are recommended for characterizing the colors of extended objects, since
the light is measured consistently through the same aperture in all bands, while
the ``cmodel' magnitudes provide a more reliable estimate of the total flux
from the galaxy that accounts for the effects of local seeing.

Before we present science results using PCA-based masses, 
we compare them with the photometrically-derived ones.  
For the DR7 sample, photometric masses are given in the
MPA/JHU catalog and derived from the 
$u, g, r, i, z$ broad-band photometry as described in \S2.1.
For the BOSS sample, we do not have an independent set 
of photometrically-derived  stellar masses for comparison. However,
we estimate the stellar mass-to-light ratio $M_*(g,r,i,z)/L_i$ 
by fitting the observed $g, r, i, z$-band fiber fluxes 
using the same model library.  We have not included the $u$-band 
in our fits, because the CMASS galaxies are 
faint in the $u$-band and  the errors are large.  
In order to avoid any uncertainties 
due to aperture effects, we use fiber masses in this whole section.

\begin{figure}
\bc
\hspace{-0.6cm}
\resizebox{8.5cm}{!}{\includegraphics{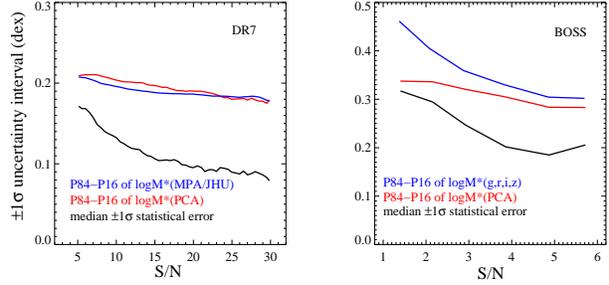}}\\%
\caption{The $\pm 1\sigma$ uncertainty interval in the logarithm of the stellar mass  
for DR7 (left) and CMASS (right) galaxies. 
Results are plotted  as a function of the  median S/N per pixel
in the spectrum. In both panels,  red lines show the median value of 
P84 $-$ P16 as a function of S/N;  P84 and P16 are 
the 84th and 16th percentiles of the cumulative PDF 
of the  PCA-based stellar mass estimate ${\rm log} M_* ({\rm PCA})$ for
each galaxy.  
Blue lines show the same quantity for the photometrically-derived 
stellar masses.
The black lines are the median $\pm1\sigma$ statistical errors derived from repeat observations.
\label{mass_err}}
\ec
\end{figure}

\begin{figure}
\bc
\hspace{-0.6cm}
\resizebox{8.5cm}{!}{\includegraphics{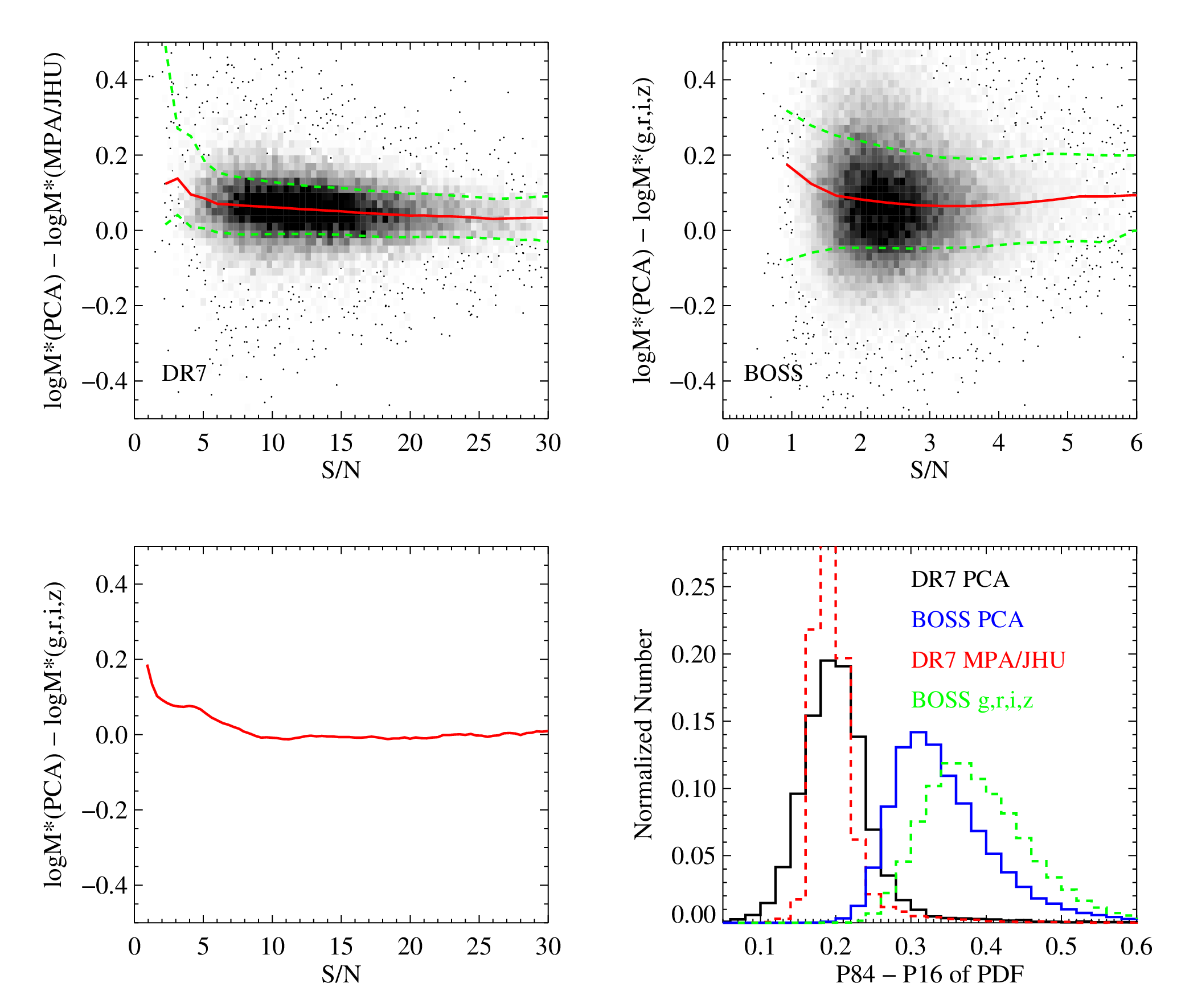}}\\%
\caption{The difference between our PCA-based stellar masses and those
derived from broad-band photometry is plotted 
as a function of  median S/N per pixel for DR7 galaxies (top-left)
and for the CMASS sample (top-right). In both panels, the over-plotted red line is 
the median, the two green dashed lines show the 68\% spread.
Bottom-left: the median discrepancy between PCA-based stellar masses and those derived from $g, r, i, z$-band photometry 
as a function of spectral median S/N per pixel for the combined DR7 and CMASS samples.
Bottom-right:  histograms of the distribution of (P86 $-$ P14), 
where P84 and P16 are the 84th and 16th percentile points of the cumulative PDF
of the  stellar mass estimated (derived using PCs)  for DR7 galaxies (black) and for
CMASS galaxies (blue). 
Red and green dashed lines show histograms of the same quantity
for stellar masses derived from photometry for  DR7 and BOSS, respectively.
\label{comp_mass}}
\ec
\end{figure}

Figure \ref{mass_err} shows the $\pm 1\sigma$ uncertainty interval in $\log M_*$ 
for DR7 (left) and CMASS (right) galaxies. 
Results are plotted  as a function of the  median S/N per pixel
in the spectrum. In both panels, the red lines show the median values of
P84 $-$ P16 in each S/N bin; P84 and P16 are 
the 84th and 16th percentiles of the cumulative PDF 
of  our PCA-based stellar mass estimate ${\rm log} M_* ({\rm PCA})$.
The blue lines track the same quantity for the photometrically-derived stellar mass
estimates.
The black lines represent the  $\pm1\sigma$ scatter in 
the $\log M_*$ estimates derived from repeat observations.
For DR7 galaxies, the errors on the stellar mass are virtually identical when using 
PCA to fit the spectra, or when fitting to the photometry.   
For BOSS, the PCA method
yields significantly smaller errors, particularly at low
S/N. This behaviour is expected, because the BOSS galaxies with low S/N spectra are
faint galaxies where photometric errors tend to be large. 

We also note that  P84 $-$ P16 of ${\rm log} M_* ({\rm PCA})$
is in general {\em larger} than the median $\pm1\sigma$ statistical
scatter in $\log M_*$ derived from repeat observations. 
This feature arises because the latter is only a measure of the error on $\log M_*$
due to noise in the spectra; the former accounts for  both 
noise and the fact that different model galaxies with different $M_*/L$ values occupy
the same region of PC-space.

\subsection{Comparison with photometrically-derived stellar masses}
In Figure \ref{comp_mass}, we  
compare the PCA-based stellar masses with the photometric masses. 
The top-left panel of Figure \ref{comp_mass} shows the difference between
${\rm log} M_*({\rm PCA})$ and ${\rm log} M_*({\rm MPA/JHU})$ 
as a function of S/N for DR7 galaxies.
The red line is the median, the two green dashed lines show the 68 percentile spread.
There is a systematic $\sim$0.05 dex offset between the two mass estimates,
but the scatter is quite small ($< \pm 0.1$ dex at S/N $>10$).
The offset of our PC-based stellar mass estimates to slightly higher
values is expected given that our model library includes
 truncated SFHs and a smaller fraction of galaxies with recent bursts  (10\% instead of  
50\%). In the next section, we will discuss
how different assumptions about the mixture of  SFHs in the model library influence our 
stellar mass estimates.

The top-right panel of  Figure \ref{comp_mass}  shows the difference between 
${\rm log} M_*({\rm PCA})$ and ${\rm log} M_*(g,r,i,z)$
for CMASS galaxies. Although these two 
sets of stellar masses are derived using exactly the 
same model library, the PCA-based stellar masses
are typically 0.08 dex higher than the ones estimated 
using the $g, r, i, z$-band photometry.  
In the  bottom-left panel of this figure, we plot the median 
discrepancy as a function of spectral median S/N per pixel for the combined DR7 and CMASS samples, finding that the offset disappears   
when  S/N $>$ 8. We conclude that in the limit of high S/N, the two
methods of estimating stellar mass yield exactly the same results. 

Finally, the bottom-right panel of  Figure \ref{comp_mass} shows 
the distribution of the $\pm 1\sigma$ errors (i.e. P84 $-$ P16)
for our  different sets of stellar mass
estimates.  Black and blue histograms are for the PC-based
stellar masses for DR7 and CMASS. The 
dashed red and green histograms show error distributions for
the stellar masses derived from the photometry.
For DR7 galaxies, the $\pm1\sigma$ uncertainties 
on the  PCA-based and photometry-based masses peak at nearly the 
same value, with the DR7 photometric measurements 
having slightly less dispersion. However, for the CMASS sample,  
the errors on the photometry-based masses are on average $\sim$0.05 dex 
larger than those derived using Principal Components, again reflecting the fact
that the photometric errors are larger for this sample.  

\subsection{Dependence of our stellar mass estimates on the input
parameters of the model library}

In this section, we study the sensitivity of our stellar mass estimates
to  the input parameters of the model library.
We change the input SFHs, dust extinction values and metallicity distributions in
the model library one at a time, and quantify the effect on the stellar masses.  

\begin{figure}
\bc
\hspace{-0.6cm}
\resizebox{6.5cm}{!}{\includegraphics{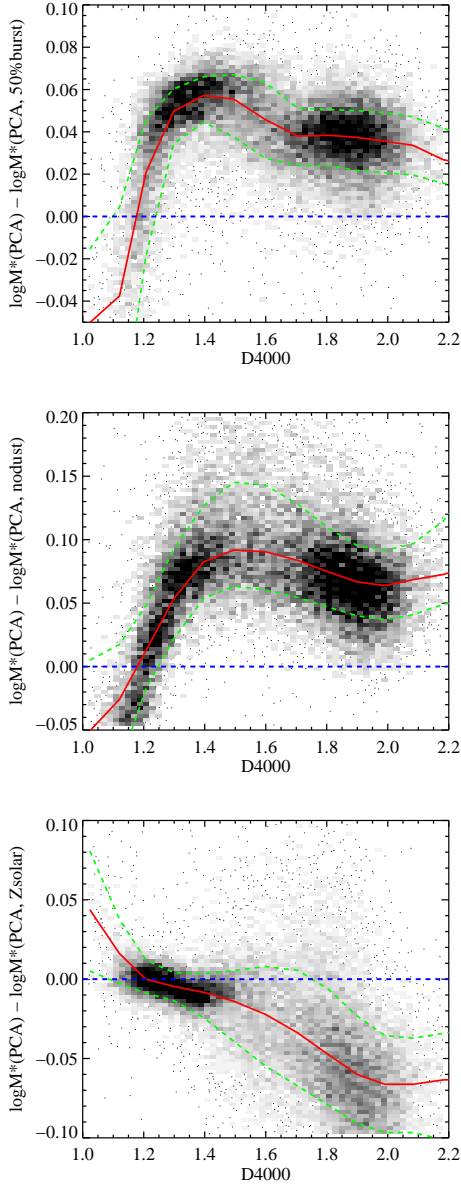}}\\%
\caption{The dependence of stellar masses on the input parameters of the model library.
Top panel: the difference between stellar masses based on the standard library and a library  
with a 50\% burst fraction; middle panel:  the difference 
between stellar masses based on the standard library and a library with no dust extinction; 
bottom panel: the difference between stellar masses based on the standard library and  
a solar metallicity library. In all these plots, 
the red lines denote  the median of $\Delta {\rm log} M_*$, 
and the two green dashed lines show the 68\% spread.
\label{Mdepend}}
\ec
\end{figure}

\subsubsection{Star formation histories}
In the previous section, we explained the 0.05 dex systematic 
difference between $M_*$(PCA) and $M_*$(MPA/JHU) as a consequence of the 
different SFHs used in generating the libraries. To confirm this idea,   
we have generated a new set of stellar mass estimates using a library in which
the burst fraction is increased from 10\% to 50\%.  
The top panel of Figure \ref{Mdepend} shows the difference in the stellar mass estimates
$\Delta {\rm log} M_* = {\rm log} M_*({\rm PCA}) - {\rm log} M_*({\rm PCA, 50\% burst})$
as a function of D4000, where $M_*({\rm PCA, 50\% burst})$ is the library with 50\% burst  
fraction.
Once again the red line in the top panel of Figure \ref{Mdepend} shows the median value of 
$\Delta {\rm log} M_*$, while the two green dashed lines show the 68\% spread. 
In the range of $1.4 < D4000 < 2.2$,
$\Delta {\rm log} M_* \sim 0.05$, consistent with the systematic offset  
found between our PC-based mass estimates and the photometrically derived
ones for DR7 sample in \S5.1. At lower values   
of  D4000, galaxies are constrained to have formed a significant fraction of
their stars recently, so the difference in the fraction of bursty galaxies
in the library makes a much smaller difference to the results.

\subsubsection{Dust extinction}
To check how assumptions about dust extinction influence our stellar mass
estimates, we generate a library without dust extinction.
The middle panel of Figure \ref{Mdepend} shows the difference 
$\Delta {\rm log} M_* = {\rm log} M_*({\rm PCA}) - {\rm log} M_*({\rm PCA, nodust})$ 
as a function of D4000.  
$\Delta {\rm log} M_*$ increases as a function of D4000 up to value of $\sim 0.08$
at  D4000 $\sim 1.5$,
and then remains approximately constant.   
The systematically smaller stellar mass-to-light ratios derived using
the library with no dust extinction can be  understood,  
because a smaller fraction of the optical light from the galaxy
is assumed to be absorbed.  
If dust is not included in the models and D4000 is large,  the fit is able to 
re-adjust to match with an older stellar population, which has
higher mass-to-light ratio. At low D4000 values, the stellar population
is more tightly  constrained to be young, so the degeneracy is again less important.   

\subsubsection{Metallicity}
To check how assumptions about metallicity influence our stellar mass
estimates, we generate a library with only solar metallicity models.
The bottom panel of Figure \ref{Mdepend} show the difference  
$\Delta {\rm log} M_* = {\rm log} M_*({\rm PCA}) - {\rm log} M_*({\rm PCA, Zsolar})$
as a function of D4000.  
As can be seen, for the young populations (D4000 $<$1.5) the systematic effects induced by adopting incorrect metallicity     
assumptions are very small. 
For the older populations, the difference increases with D4000, which
is indicative of the well-known  age-metallicity degeneracy. 

In summary, stellar mass estimates are most strongly affected by
assumptions about dust extinction, but also by the 
SFHs and metallicity of the model library galaxies.
In all  three cases,  
systematic offsets are of order  0.05$-$0.1 dex in $\log M_*$. Note
we have not considered changes to the IMF.
To convert our \citet{kroupa01} stellar masses to a
\citet{salpeter55} or  \citet{chabrier03} IMF, one should add 0.18 dex to or subtract
0.05 dex from  the logarithm of stellar masses, respectively.
We note, however, that the \citet{salpeter55} IMF is disfavoured by dynamical $M_*/L$ estimates of
elliptical galaxies \citep{cappellari06}.
 
\subsection{The dependence of stellar masses on the assumed 
stellar population synthesis model}

\begin{figure}
\bc
\hspace{-0.6cm}
\resizebox{8.5cm}{!}{\includegraphics{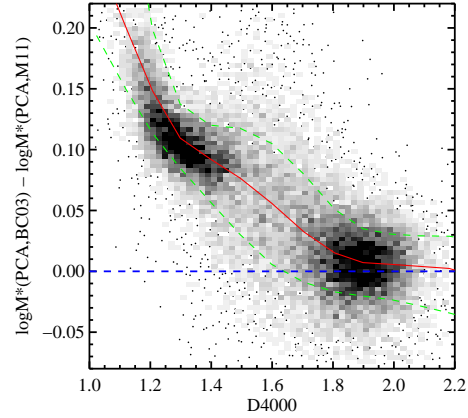}}\\%
\caption{The difference in stellar masses derived from
solar metallicity libraries based on BC03 and M11 models.
\label{model_depend}}
\ec
\end{figure}

\citet{kannappan07} compared photometrically-derived stellar masses 
based on BC03 and \citet{maraston05} population synthesis models. 
These models are quite different in the optical due to the use 
of different stellar libraries.
They found that  BC03 models yield stellar mass estimates that are $\sim$1.3 times  
larger, even when no near-IR photometry is used in the fits. 

In this section, we investigate systematic uncertainties that may arise in our PCA-based 
stellar mass estimates as a result of our  choice of  stellar population synthesis model.
Maraston \& Stromback (2011) present high spectral resolution stellar population
models using the 
MILES stellar library \citep{miles06} as input.
The models have been extended with SSPs based on theoretical stellar
libraries, which extend the wavelength coverage of the models into the
ultra-violet.

We  compare results at solar metallicity, where   
the stellar age coverage of the current version of the M11 model we are using is
most reliable.
Figure \ref{model_depend} shows the 
difference in the stellar masses estimated using these two libraries 
as a function of D4000. 
The strongest systematic discrepancy appears at young ages
(low values of D4000),  and decreases for older stellar populations.  
The offset of $\sim 0.12$,
consistent with the result of \citet{kannappan07}, arises because the M11 models
use Geneva tracks \citep{schaller92, meynet94} to model stellar evolution, while the BC03 models 
make use of Padova tracks \citep{alongi93, bressan93, fagotto94a, fagotto94b, girardi96}.
In these tracks different assumptions are made
regarding convective overshooting and the 
temperature and energetics of the Red
Supergiant Phase, leading to a significantly
redder supergiant phase in the M11 models. 

We note that the prescription for stellar mass loss  
is somewhat different betwimeen BC03-models
and Maraston-models, so we 
have used the BC03 mass loss prescriptions
when performing the comparison shown in Figure 14.
 
\section{ Evolution of Massive Galaxies to Redshift  0.6}

\begin{figure}
\bc
\hspace{-0.6cm}
\resizebox{8.5cm}{!}{\includegraphics{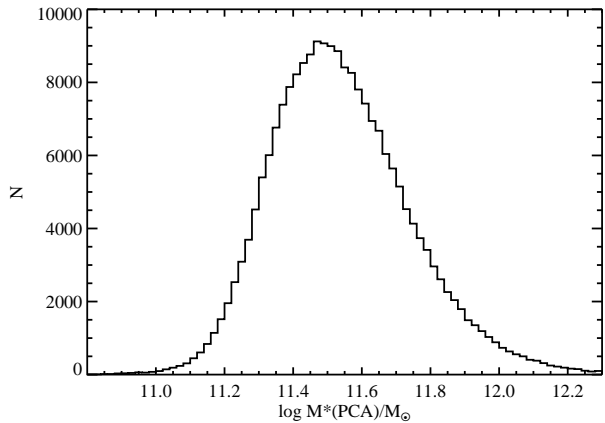}}\\%
\caption{The distribution of our PCA-based stellar masses for BOSS CMASS sample.
\label{boss_mass_dist}}
\ec
\end{figure}

Figure \ref{boss_mass_dist} shows the distribution of total stellar
masses based on the PCA method for the BOSS CMASS sample\footnote{In
  this section, we use total stellar masses for both DR7 and BOSS
  samples. When we use the PCA method to estimate the total stellar
  mass, the underlying assumption is that the stellar mass-to-light
  ratio within the fiber aperture is the same as that for the entire
  galaxy.}.  As can be seen, the range of stellar masses spanned by
this sample is rather narrow: $11 \le \log M_* \le 12$. In the local
Universe, galaxies in this stellar mass range are mainly ``red and
dead'' systems with no ongoing star formation.  Many reside in groups
and clusters and have radio jets, which are believed to heat the
surrounding gas, preventing it from cooling and forming stars
\citep{boehringer93, mcnamara00, fabian03, best05a, best05b, best06,
  bower06, croton06}. According to the currently popular
``down-sizing'' scenario for galaxy evolution, galaxies of this mass
should have completed their star formation at very early epochs
\citep[$z>2$;][]{heavens04, thomas05}. They may subsequently grow
through merging, but these merger events are believed to be largely
dissipationless \citep[i.e. ``gas-free";][]{naab06, bell06,
  scarlata07, kang07}.  If this picture is correct, we would expect
the stellar populations of very massive galaxies to evolve only
``passively'' over the redshift interval from 0.6 to 0. In this
section, we check whether these expectations are correct.

\subsection{Fraction of massive galaxies with young stars }

\begin{figure*}
\bc
\includegraphics[angle=0,width=0.33\textwidth]{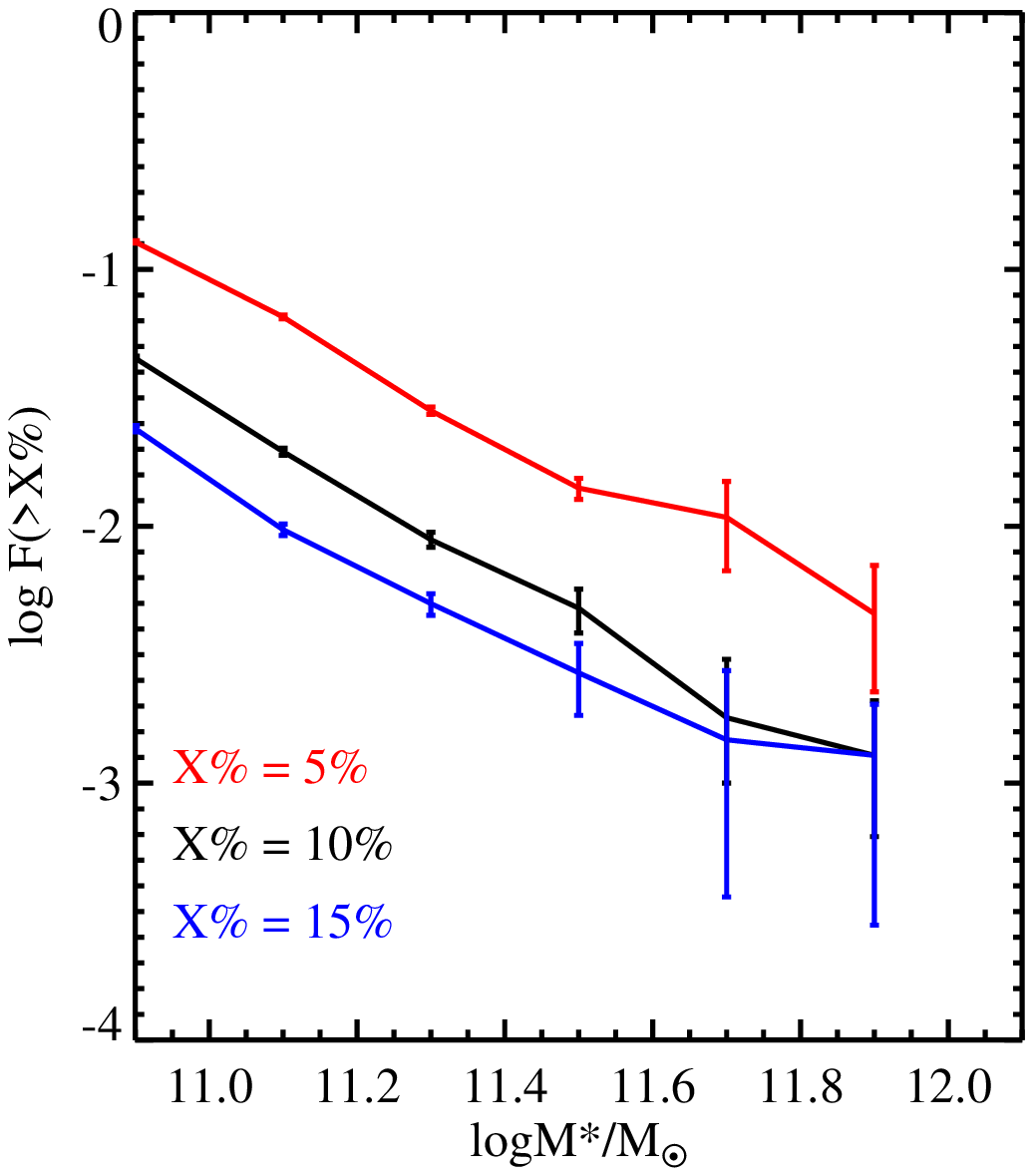}
\includegraphics[angle=0,width=0.33\textwidth]{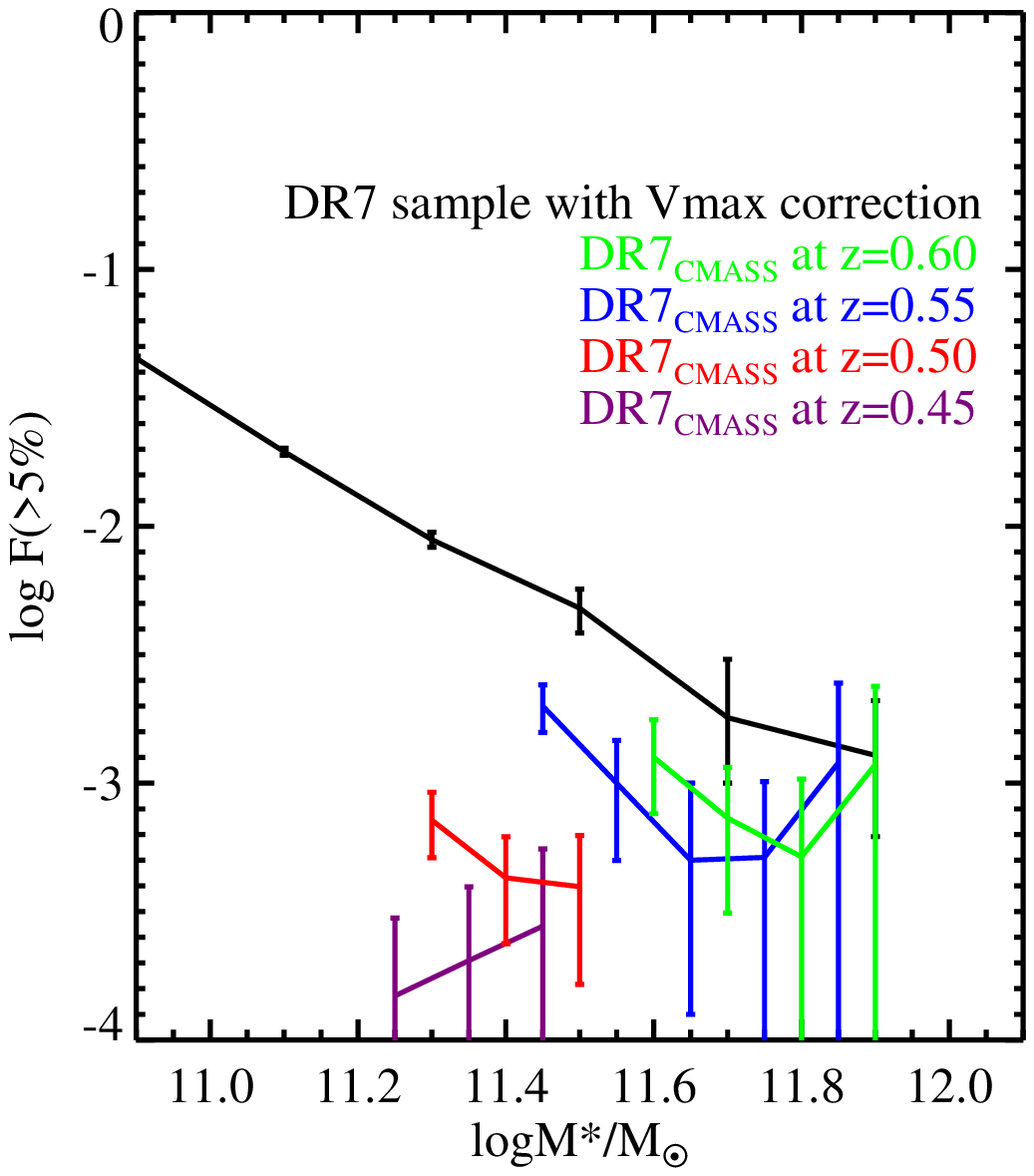}
\includegraphics[angle=0,width=0.33\textwidth]{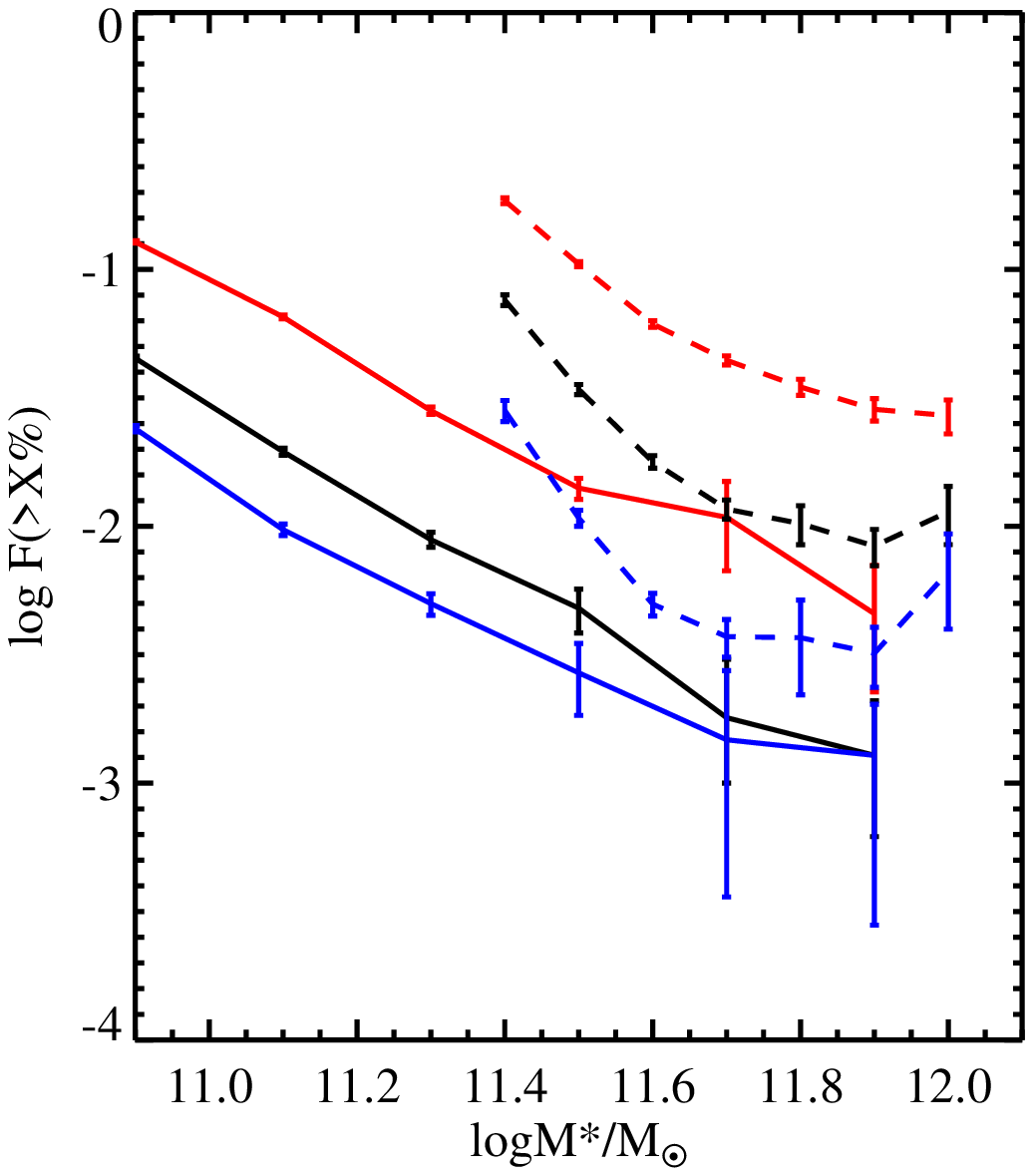}
\caption{This figure shows the fraction of galaxies which have 
$>$ X\% stars formed in the last Gyr as a function of 
stellar mass. Left panel: The red, black and blue lines show ${\rm log} F(>5\%)$, ${\rm log} F(>10\%)$
and  ${\rm log} F(>15\%)$ for the DR7 sample. Middle panel:
The purple, red, blue, green lines (they have been slightly displaced
in the x-axis by minus 0.1, 0.05, 0.0, $-$0.05dex, respectively) are results for the redshifted DR7 samples 
at $z = 0.45,0.5,0.55$ and 0.6, after application of the CMASS color cuts, compared to the results obtained for
the unredshifted DR7 sample (in black). Right panel: The dashed lines
show log$F(>5\%)$,  log$F(>10\%)$ and  log$F(>15\%)$ for CMASS
sample with $z > 0.54$ and  $\log M_* > 11.4$; this is a lower limit since the $d_\perp>0.55$ constraint
deletes some blue galaxies from the sample. Solid lines show the DR7 results for reference.
\label{f5per}}
\ec
\end{figure*}

Our PC-decomposition technique can isolate galaxies which have 
had more than a few  percent of their stellar masses
formed in the last Gyr (see panel (h) of Figure \ref{coef_cx}).
We explore the robustness of the SFHs derived by the PCA method in the Appendix.
In this section, we define parameters $F(>5\%)$, $F(>10\%)$, and $F(>15\%)$, which
represent the fraction of galaxies in which more than 5, 10 and 15\%
of the stellar mass was formed in the last Gyr, respectively.  Using
DR7 and BOSS data, we study how these fractions have evolved since $z
\sim 0.6$ for galaxies more massive than $\sim 2 \times 10^{11}
M_{\odot}$.

Starting from the DR7 Main galaxy sample, we construct a magnitude-limited 
sample of galaxies with  $14.5 < r < 17.6$ and
redshifts in the range $0.055 < z<0.3$. We also limit the sample to galaxies 
with ZWARNING = 0 and SPECPRIMARY =1 to eliminate   
repeat observations and potential redshift errors. These restrictions result in a final DR7 sample of $\sim$430,000 galaxies. 
For each observed galaxy $i$, we define the quantity $z_{{\rm min},i}$ 
and  $z_{{\rm max},i}$ to be the
minimum and maximum redshift at which the galaxy would satisfy the
apparent $r$-band magnitude limit. Evolutionary
and K-corrections are included in this calculation  
\citep{li07,li09}.
 This allows us to define $V_{{\rm max},i}$ for the galaxy as the total 
comoving volume of the survey between $z1$ and $z2$, 
where $z1$ is the maximum of $z_{{\rm min},i}$ and 0.055, and 
$z2$ is the minimum of $z_{{\rm max},i}$ and $0.3$. 
$F_*>X\%$ can then be estimated as
\begin{equation}
F(>X\%) = \frac {\sum_{i = 1,N_{\rm act}}(V_{{\rm max},i})^{-1}}{\sum_{j = 1,N_{\rm all}}(V_{{\rm max},j})^{-1}},
\end{equation}  
where the sum on the numerator extends over  $N_{\rm act}$,
the number of galaxies in 
a given stellar mass bin that have formed more than X\% of their
stars in the last Gyr, while the sum on the 
denominator extends over $N_{\rm all}$, the total number
of galaxies in the same mass bin. 

The red, black and blue lines in the left panel of 
Figure \ref{f5per} show ${\rm log} F(>5\%)$, ${\rm log} F(>10\%)$
and  ${\rm log} F(>15\%)$ as a function of stellar mass for 
the DR7 sample (errors are derived from boot-strapping). 
As can be seen, all three  fractions 
decrease strongly and monotonically with 
increasing stellar mass. At all stellar masses, there are 10
times more galaxies that have formed more than 5\% of their stars over the last
Gyr, compared to the number that have formed more than 15\% of their stars
in the last Gyr. 

Before comparing these results with corresponding values of $F$
for CMASS galaxies at $z\sim 0.6$, 
we note that CMASS sample is not a simple magnitude-limited
sample. There is 
a $d_\perp>0.55$ color cut, which means that  blue galaxies  will be lost from the
survey, particularly at the lower redshift end. This means that the fraction of
actively star-forming galaxies that we compute for the CMASS galaxies 
represents a {\em lower limit} to the true value. 

Because we do not know the underlying relation between color and stellar mass 
for galaxies with $M_*> 2\times 10^{11} M_{\odot}$  
at $z \sim 0.55$ (existing surveys do not extend over wide enough areas to sample 
large numbers of very massive galaxies), it is difficult to 
correct for any missing blue galaxies.   
In order to provide a more quantitative idea of the degree to which the $d_\perp>0.55$ cut
{\em might} affect our estimates of the fraction of galaxies with  recent star formation, 
we use the K-correct code 
\citep{blanton07} to predict the colors of galaxies in the DR7 sample at
redshifts  $z = 0.45, 0.5, 0.55$ and $0.6$.  
At each redshift, we  select objects that pass the 
CMASS target selection criteria. In addition, we define a
redshift-dependent lower stellar mass limit  
${\rm log} M_*^{\rm lim}/M_\odot = 2.0 \times z + 10.35$, so that  
a passively-evolving galaxy  at redshift $z$ with 
stellar mass $M_* > M_*^{\rm lim}$  would pass
all the target selection criteria in equation (1)
(with the exception of  $i_{\rm fiber2} < 21.5$, which is more
difficult to estimate unless one has a model for the structure of the galaxy). 

The colored lines in the middle  panel of  Figure \ref{f5per} 
show ${\rm log} F(>10\%) = N_{\rm act} / N_{\rm all}$ for 
the four redshifted DR7 samples (the purple, red, blue, green lines have been slightly displaced
in the x-axis by minus 0.1, 0.05, 0.0, $-$0.05dex, respectively), compared to the results obtained for
the unredshifted DR7 sample (in black) .
As can be seen, the fraction of  massive galaxies with 
recent star formation could  be under-estimated by
more than a order of magnitude at $z<0.55$. At higher redshifts, the fraction
of actively star-forming galaxies that is missed is closer
to a factor $\sim 2$. 
We therefore select a sub-sample of the CMASS galaxies 
with $ 0.54 < z < 0.7$ (0.54 is the  
median redshift of the CMASS sample) and log$M_*/M_\odot > 11.4$
as the main high-redshift  comparison sample for the DR7 massive galaxies.

In the right panel of Figure \ref{f5per}, the dashed lines
show log$F(>5\%)$,  log$F(>10\%)$ and  log$F(>15\%)$ as a function of stellar mass for this
sample. Solid lines show the DR7 results for
reference.  $z_{{\rm min},i}$ and $z_{{\rm max},i}$ values have been calculated for this sample
by evaluating the     
minimum and maximum redshifts at which the galaxy would satisfy 
all the criteria in equation (1) except $i_{\rm fiber2} < 21.5$. Evolutionary
and K-corrections are included in this calculation. 
$V_{{\rm max},i}$ is calculated as the 
comoving volume of the survey between $z1$ and $z2$, 
where $z1$ is the maximum of $z_{{\rm min},i}$ and 0.54, and 
$z2$ is the minimum of $z_{{\rm max},i}$ and $0.7$.

\begin{figure}
\bc
\hspace{-0.6cm}
\resizebox{8.5cm}{!}{\includegraphics{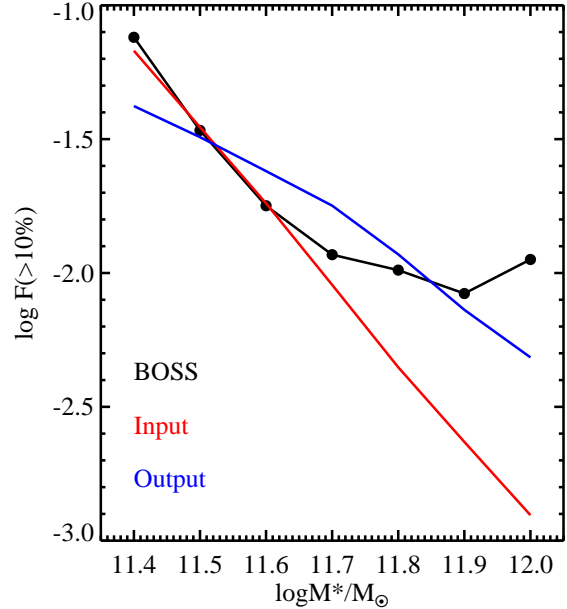}}\\%
\caption{The black line  
reproduces  the black dashed line in the right panel of Figure \ref{f5per}. 
The three low mass data points on this black line are fitted with a linear relation 
${\rm log} F_*(>10\%) = -3 \times {\rm log} M_*/M_\odot + 33$. 
We randomly generate a sample of 1,000,000 galaxies 
with ${\rm log} M_*/M_\odot$ has a Gaussian distribution over the range of [11.1, 12.1]
with a peak at $\sim$11.55 and 68\% of them distributed 
over the range 11.37$-$11.73. We assign a  value of $F_*$  
to each galaxy so that the linear relation is reproduced on average (red line
on the plot). We then add an error to each stellar mass using the error distribution of    
CMASS galaxies as a function of $M_*$. This mimic sample has a similar distribution 
in ${\rm log} M_*/M_\odot$ as our CMASS galaxies within $0.54 < z < 0.7$. We then recompute
the relation between ${\rm log} F_*(>10\%)$ and stellar mass,  the result is shown as 
the blue line.
\label{mimic}}
\ec
\end{figure}

One conclusion from Figure \ref{f5per} is that the fraction of actively star-forming
galaxies with $\log M_* > 11.4$ has evolved strongly since
a redshift of $\sim 0.6$. This result is not surprising. The evolution of the 
dependence of star formation on galaxy stellar mass has been studied using
data from other deep surveys  \citep[e.g.,][]{zheng07, chen09, karim11}
and, in general, the claim was been that the rate of decline in cosmic SFR
with redshift is the same at all stellar masses. None of these previous
surveys, however, have extended to stellar masses as high as $10^{12} M_{\odot}$. 
In the BOSS data, we find  the striking result that the fraction of
actively star formation galaxies {\em flattens} above a stellar mass of
$10^{11.6} M_{\odot}$ at $z\sim 0.6$.
At the largest stellar masses, therefore, the evolution in the fraction of 
star-forming galaxies from $z \sim 0.6$ to the present-day is even 
more dramatic, reaching a factor of $\sim 10$ 
at $\log M_* \sim 12$.  We emphasize once again that these numbers
represent {\em lower limits} on the evolution, because of the 
incompleteness issues described above. 

Because the uncertainties in the BOSS stellar 
masses are larger than our bin size, we have done tests 
to explore whether ``smearing'' of the true stellar
mass distribution could produce the observed flattening. 
In Figure \ref{mimic}, the black line  
reproduces  the black dashed line in the right panel of Figure \ref{f5per}. 
We fit the three low mass data points with a linear relation 
${\rm log} F_*(>10\%) = -3 \times {\rm log} M_*/M_\odot + 33$. 
We randomly generate a sample of 1,000,000 galaxies 
with ${\rm log} M_*/M_\odot$ has a Gaussian distribution over the range of [11.1, 12.1]
with a peak at $\sim$11.55 and 68\% of them distributed 
over the range 11.37$-$11.73. We assign a  value of $F_*$  
to each galaxy so that the linear relation is reproduced on average (red line
on the plot). We add an error to each stellar mass using the error distribution of    
CMASS galaxies as a function of $M_*$. This mimic sample has a similar distribution 
in ${\rm log} M_*/M_\odot$ as our CMASS galaxies within $0.54 < z < 0.7$. We then recompute
the relation between ${\rm log} F_*(>10\%)$ and stellar mass
and  the result is shown as 
the blue line. The main conclusion from Figure \ref{mimic} 
is that errors will act to flatten the trend, but this
is a small effect compared to what is seen in the data.  
Smearing by errors also  cannot explain the 
characteristic mass scale of $\log M_*=11.6$
where the flattening appears to set in.

\subsection{AGN and star formation in massive galaxies}

\begin{figure*}
\bc
\hspace{-0.6cm}
\resizebox{17.cm}{!}{\includegraphics{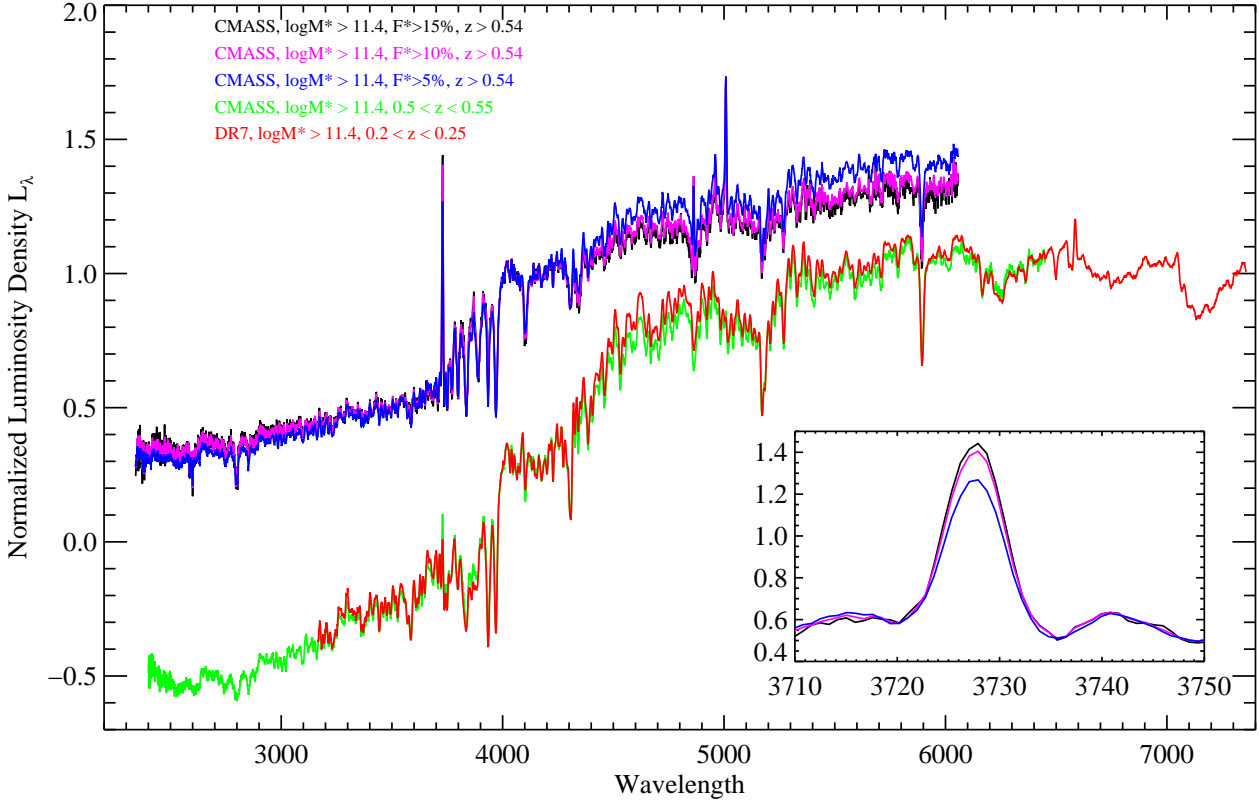}}\\%
\caption{The black, magenta and blue spectra represent the stacks of galaxies with $F_*>15\%$, $F_*>10\%$, and $F_*>5\%$ in 
CMASS sample, respectively. The red spectrum is the stack of ``inactive" ($F_*<10\%$) DR7 galaxies with  
log$M_*/M_\odot > 11.4, 0.2 < z < 0.25$, the green spectrum shows its CMASS twin sample which has exactly the same 
mass distribution and $F_*<10\%, 0.5 < z < 0.55$. The insert drawing highlights the region near the \oii\ line.
\label{stack_spec}}
\ec
\end{figure*}

The CMASS galaxies are massive (log$M_*/M_\odot > 10^{11.2}$) and
predominantly bulge-dominated \citep{masters11},
therefore it is likely that they host supermassive black holes.
In bulge-dominated galaxies where star-formation is
ongoing, the black holes are usually accreting actively \citep{heckman04, kauffmann09}.  We
now turn to examining the black hole accretion rate using the \oiii
$\lambda5007$ line as an indicator of the AGN's bolometric
luminosity.   Because the individual spectra are noisy, we average
them to create high-S/N composites.

We present stacked spectra of CMASS galaxies 
with  $F_*>X\%$ ($X\%$ = 5\%, 10\% or 15\%), log$M_*/M_\odot > 11.4, z > 0.54$.
The galaxy spectra are first corrected for foreground Galactic
attenuation using the dust maps of \citet{schlegel98},
transformed from vacuum wavelengths to air, from flux densities 
to luminosity densities,  and shifted to the
rest frame using the redshift determined by the BOSS
pipeline. The rest-frame spectra are averaged with weight $1/V_{\rm max}$
(Note that the weight of the bad pixels
identified in the SDSS mask array is set to zero). Finally, we 
normalize the stacked spectra by their mean luminosity in
the wavelength range 4000$-$4080\AA.
The black, magenta and blue 
spectra in Figure~\ref{stack_spec} represent the normalized
stacks of CMASS galaxies with  $F_*>15\%$, $F_*>10\%$, and $F_*>5\%$, respectively. 
As expected, the \oii\ flux increases for larger values of $X$ and the
spectra are also bluer.

In the following, we will concentrate on  the stack with $F_*>10\%$.
In order to quantify how  AGN contribute to the line emission, 
we have selected DR7 galaxies with similar values of D4000, H$\delta_A$, log(\oiii/H$\beta$),
log(\oii/H$\beta$). When we combine the spectra of these ``matched'' DR7 galaxies,
we find  log(\nii/H$\alpha$) $\approx -0.5$.   
log(\oiii/H$\beta$) is measured directly from the CMASS stack
and is  $\approx 0.2$. This implies that around half of the \oiii\ luminosity in the 
stack is contributed by AGNs \citep{kauffmann03b, kauffmann09}. 
We fit the stacked CMASS spectrum as a non-negative linear 
combination of single stellar population models,
with dust attenuation modeled as an additional free parameter \citep{brinchmann04, tremonti04}.
This fit yields  a continuum $V$-band dust extinction value of 1.76.
Taking a mean bolometric correction to the extinction-corrected \oiii\ of 600 \citep{kauffmann09}, 
and assuming that  half of the \oiii\ emission 
is coming from the AGNs, we find $L/L_{\rm Edd} \approx 0.01$, where
$L_{\rm Edd} = 1.38 \times 10^{38}M_{\rm BH}/M_\odot$ is the Eddington 
luminosity and $M_{\rm BH}$ is 
estimated from the median value of the velocity dispersions of the galaxies that go 
into the stack using the formula given in  \citet{graham11} \footnote{\citet{graham11} suggests a slope of 5 rather than 4
\citep{tremaine02, graham08, graham09} for the stellar mass $-$ stellar velocity dispersion relation, which is 
also what \citet{hu08} found when considering only the massive
galaxies.}.

We have also cross-matched the BOSS and FIRST surveys, and found
  that $\sim$2.4\%CMASS galaxies have FIRST detections. The typical 
  $i$-band magnitude and mass of this radio loud sample are 19.6 and $10^{11.6}M_\odot$, respectively.
For this radio-detected sample, we construct a 
control sample matched in redshift, stellar mass, and velocity
dispersion which are located in the FIRST survey area
but lack radio detections. Interestingly, the 
fraction of galaxies with recent star formation 
($F_*>10\%$) is 2$-$2.5 times {\em smaller} in the radio-loud sub-sample than for the controls.
We will study the difference between radio-loud and radio-quiet CMASS galaxies in more detail
in future work. 

The red spectrum in Figure~\ref{stack_spec} is a stack of
DR7 galaxies with $F_*<10\%$, log$M_*/M_\odot > 11.4$ and $0.2 < z < 0.25$. 
In order to make a fair comparison 
between DR7 and BOSS ``inactive" galaxies, we construct a twin sample from CMASS
galaxies in the redshift range $0.5-0.55$, which  
has exactly the same stellar mass distribution and $F_*<10\%$.
The stacked spectrum of this twin sample is shown in green. 
(Note that the red and green spectra have both been shifted down by 0.7 from the other three spectra. The stacks are generated 
in the same way as the ``active" star forming galaxies except we use an equal weight rather than $1/V_{\rm max}$). 
There is no apparent difference in the spectral shape and absorption 
line features of the stacks of DR7 and BOSS 
``inactive" galaxies.    

In summary, we conclude that at $z\sim0.6$ at least 2\% of very massive galaxies 
have formed more than 10\% of their stars in the last Gyr and nearly 10\%
have formed more than 5\% of their stars over this period.
We note that for a galaxy 
with $M_* \sim 10^{12} M_{\odot}$ to be counted as a member of the  $F* >10\%$
``class'', it must have processed more than $10^{11} M_{\odot}$   
of gas over the last Gyr or so, i.e.  6$-$8 times as much gas
as contained in the Milky Way! If this gas is in molecular form, it should
be easily detectable at $z \sim 0.6$. 
More detailed studies 
of these objects will  reveal important insights into the physical 
processes that govern the evolution of massive galaxies at late times.
\citet{kavira08}  studied recent star formation 
in massive galaxies at $z \sim$ 0.6 using rest-frame UV data, and found 
 young stars  at levels of a few percent by mass fraction. 
Based on a strong correspondence between 
the presence of star formation (traced by UV colours) 
and the presence of morphological disturbances, 
\citet{kavira11} suggested the star formation is merger-driven. 
The major merger rate at late epochs 
($z<1$) is predicted to be  too low to produce the
observed number of disturbed LRGs, so the authors invoked minor mergers
as an alternative machanism. 
Future kinematic studies of larger sample 
of such galaxies would help test this hypothesis.

\section*{acknowledgements}
We thank the anonymous referee for suggestions that led to improvements in this
paper. The research is supported by the National Natural Science Foundation of China (NSFC) under
NSFC-10878010, 10633040, 11003007 and 11133001, the National Basic Research Program (973 program
No. 2007CB815405)  and the National Science Foundation of the United
States Grant No. 0907839.  Funding for the SDSS-I/II has been provided by the Alfred P. Sloan Foundation, the 
Participating Institutions, the National Aeronautics and Space Administration, the National Science
Foundation, the U.S. Department of Energy, the Japanese Monbukagakusho, and the Max Planck Society. 
The SDSS Web site is http://www.sdss.org/.  The SDSS is managed by the Astrophysical Research Consortium 
(ARC) for the Participating Institutions. The Participating Institutions are The University of Chicago, Fermilab,
the Institute for Advanced Study, the Japan Participation Group, The Johns Hopkins University, Los Alamos 
National Laboratory, the Max-Planck-Institute for Astronomy (MPIA), the Max-Planck-Institute for Astrophysics 
(MPA), New Mexico State University, University of Pittsburgh, Princeton University, the United States Naval 
Observatory, and the University of Washington.

Funding for SDSS-III has been provided by the Alfred P. Sloan Foundation, the Participating Institutions, the 
National Science Foundation, and the U.S. Department of Energy. SDSS-III is managed by the Astrophysical 
Research Consortium for the Participating Institutions of the SDSS-III Collaboration including the University of 
Arizona, the Brazilian Participation Group, Brookhaven National Laboratory, University of Cambridge, University 
of Florida, the French Participation Group, the German Participation Group, the Instituto de Astrofisica de Canarias, 
the Michigan State/Notre Dame/JINA Participation Group, Johns Hopkins University, Lawrence Berkeley National 
Laboratory, Max Planck Institute for Astrophysics, New Mexico State University, New York University, Ohio State 
University, Pennsylvania State University, University of Portsmouth, Princeton University, the Spanish Participation 
Group, University of Tokyo, University of Utah, Vanderbilt University, University of Virginia, University of Washington, 
and Yale University. 

\section*{Appendix}
Here we explore the robustness of the SFHs that we
infer.  Our analysis is based on restframe optical wavelengths which 
are contain a large contribution from intermediate age stars; therefore  
our code does best at recovering the SFR averaged over the last Gyr.
In contrast, commonly used SFR tracers such as H$\alpha$, the far-ultraviolet,
and the far-infrared trace star formation on timescales of
$\sim$10$-$100 Myr \citep{kennicutt98}.  

\begin{figure}
\bc
\hspace{-0.6cm}
\resizebox{8.5cm}{!}{\includegraphics{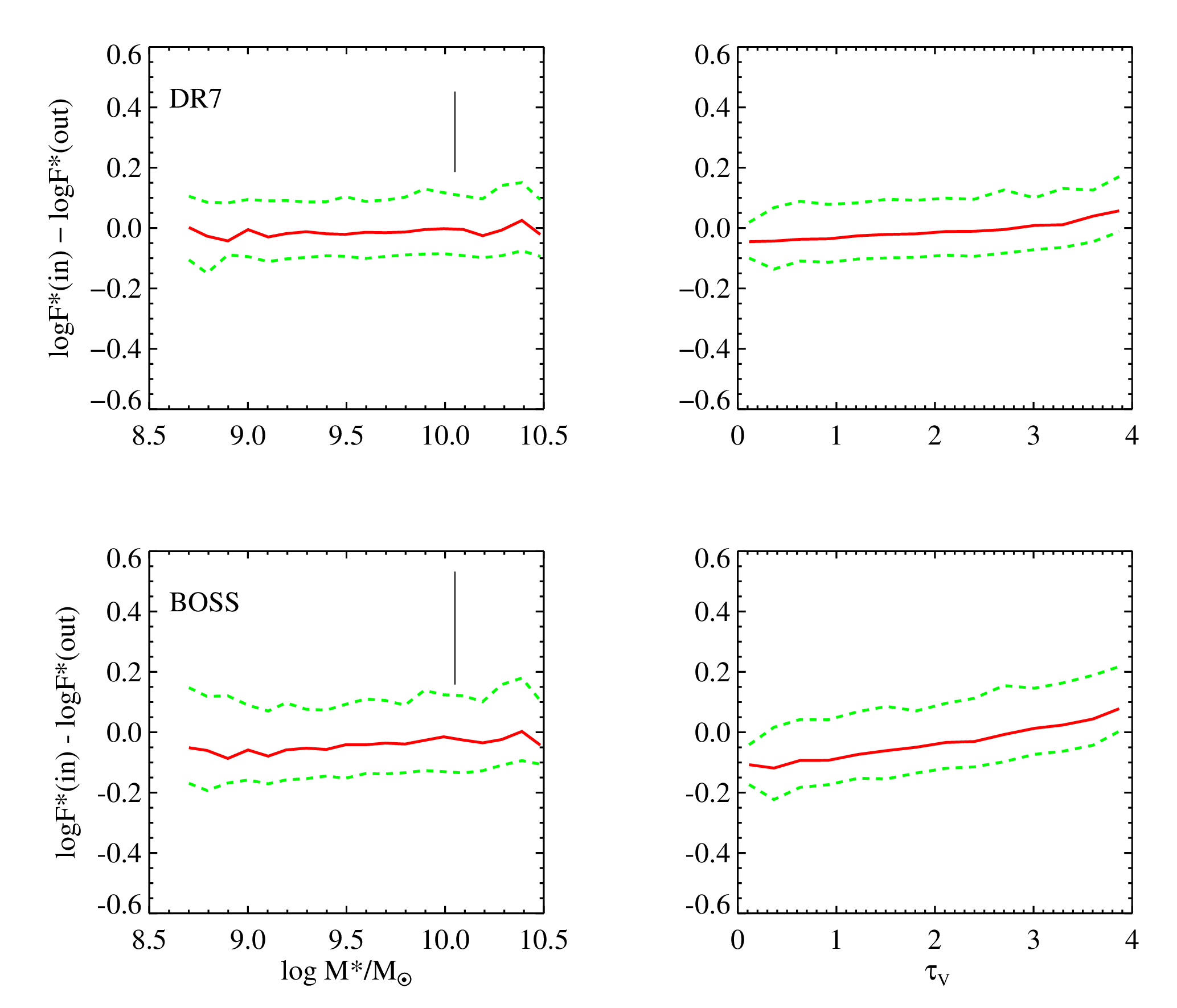}}\\%
\caption{The difference between our input, $F_*({\rm in})$, and
output, $F_*({\rm out})$ in tests using simulated data.  As described in 
$\S$6, our suite of 5000 test spectra have values of SFR,
$M_*$, $Z_*$, and $\tau_V$ drawn from DR7 star forming galaxies. The
top panels and bottom panels show simulated DR7 and BOSS data
while the left panels and right panels shows the log$F_*$
residuals versus stellar mass and $V$-band dust attenuation.
The red solid line denotes the median and the green dashed lines enclose
the 68\% of the data points.  A black error bar denotes the median $\pm 1\sigma$
error of the PCA-derived parameters.  There is good agreement between
the derived errors and the scatter in the input and output parameters, 
and only weak evidence for a systematic trend.  This suggests that the
PCA technique is relatively robust against degeneracies between age, 
dust, and metallicity and able to accurately recover the SFR in the
last Gyr when the data is well represented by the model grid. 
\label{sim_fstar}}
\ec
\end{figure}

The recovery of galaxy star formation histories from integrated
spectra is a difficult problem due to well-known degeneracies between
age, metallicity, and dust attenuation.  We explore the effect of these degeneracies  on our SFR estimates in
two ways: a) We generate synthetic data where the input parameters are
well known and test the ability of our PCA-based algorithm to recover
the true SFR. b) We use the real data to test whether our PCA-based  SFR
estimates give answers that are consistent with SFR derived from nebular
emission lines. In the real universe there
are strong correlations between SFR, stellar mass, metallicity, and
dust attenuation \citep[c.f.,][]{brinchmann04, tremonti04, gallazzi05, asari07}.  
To ensure that our suite of
test models reflected these parameter correlations, we used DR7
galaxies to define the input parameters of our model.  We randomly
selected 1000 SDSS DR7 star forming galaxies and tabulated their $M_*/L$
ratio, $V$-band dust attenuation of young stars ($\tau_V = A_V / 1.086$), nebular
metallicity, and SFR/$M_*$ from the MPA/JHU catalog. (The dust
attenuation, metallicity, and SFR have been estimated from the nebular
lines \citep{brinchmann04}.  We assume that the stellar
metallicity, $Z_*$, is 0.4 dex lower than the nebular metallicity, as
found by Gallazzi et al. 2005.)  For each galaxy we identified all the
models in our library that were within $\pm$0.1 dex in log$(M_*/L)$,
log(SFR/$M_*$), $Z_*$, and $\tau_V$, and we randomly selected 5 models
from this subset.  We used the error array of the SDSS spectrum to add
realistic random errors to each of the model spectra.  We then applied
our PCA analysis to the $\sim$5000 simulated spectra and estimated $F_*$.
In the top panel of Figure \ref{sim_fstar} we compare the input and
output values of log$F_*$ as a function of stellar mass and dust
attenuation for our simulated DR7 galaxies.  In the lower panel we
show the result of a similar exercise where we have substituted the
error arrays of randomly selected BOSS galaxies to explore our ability
to recover $F_*$ at the low S/N typical of BOSS.  For the simulated
DR7 and BOSS spectra, we recover $F_*$ to within $\pm0.1$ and $\pm0.2$
dex respectively.  There is as a small systematic trend with dust
attenuation that is evident in the noisier BOSS data, but this
produces only a very weak systematic trend with stellar mass (less
than 0.05 dex over two orders of magnitude in stellar mass).  Thus,
our PCA technique appears to accurately recover the SFR in the last
Gry in the case where the input data is well represented by the
models.

\begin{figure}
\bc
\hspace{-0.6cm}
\resizebox{8.5cm}{!}{\includegraphics{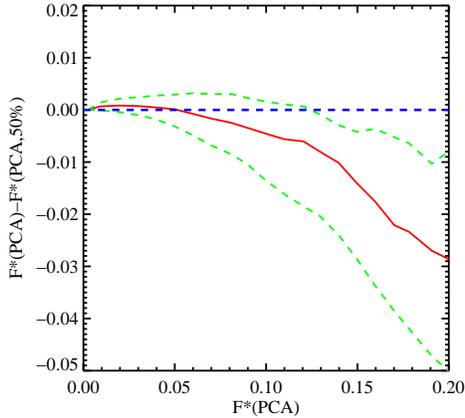}}\\%
\caption{$\Delta F_* = F_*({\rm PCA}) - F_*({\rm PCA,50\%})$ 
as a function of $F_*({\rm PCA})$,
where $F_*({\rm PCA})$ is the fraction of stars formed in the last
Gyr using the fiducial library, and  $F_*({\rm PCA,50\%})$ is the fraction 
of stars formed in the last Gyr using the library
with 50\% burst fraction. The red lines denote  the median value of $\Delta F_*$, 
and the two green dashed lines show the 68\% spread in  $\Delta F_*$. 
\label{comp_fstar}}
\ec
\end{figure}

The next question is whether our choice of priors influences the
derived value of $F_*$.  We have done a variety of tests similar to 
those outlined in \S5 and find that our derived SFHs are generally
insensitive to changes in our input model grid.  Not
surprisingly, the parameter that is most important is the fraction of  
of galaxies with bursts in the input model library. 
In Figure~\ref{comp_fstar}, we plot  
$\Delta F_* = F_*({\rm PCA}) - F_*({\rm PCA,50\%})$ as a function 
of $F_*({\rm PCA})$,
where  $F_*({\rm PCA})$ is our estimate of the fraction of young
stars formed in the last Gyr using the fiducial model library
and  $F_*({\rm PCA,50\%})$ is the fraction estimated using the library
with a 50\% burst fraction. Although the difference in the two estimates
is an increasing function of  $F_*({\rm PCA})$, in percentage terms the two
estimates give results that differ very little.  Moreover, the bulk of
our galaxies have very small values of $F_*$ where the difference
is negligible.

Finally, for the DR7 star-forming galaxies, we directly compared
the current SFR inferred from the nebular emission lines \citep{brinchmann04}
to our PCA-based estimates of the average SFR in the last
Gry (SFR $= F_* \times M_*/ 10^9$~yr).  The scatter at fixed stellar mass is ±0.2 dex.  Part of this may
stem from real differences in the SFR over the timescales probed
(10 Myr vs. 1 Gyr).
Curiously, we find systematic trends with stellar mass and dust
attenuation that are not present in our tests with simulated data
(Fig.~\ref{sim_fstar}). For the real data, the PCA technique appears to
underestimate the SFR inferred from extinction corrected H$\alpha$ by
as much as 0.4 dex in the most massive (log$(M_*/M_{\sun}) >
10.8$) or dusty ($\tau_V > 3$) galaxies.  A systematic trend of this nature is difficult to
explain by invoking star formation history differences as this would
imply that \emph{all} massive galaxies are in the midst of a burst at
the current epoch.  We note that similar discrepancies have been found
in other works that compare information inferred from the restframe
optical continuum and the nebular lines \citep[c.f.,][]{tanaka11, hoversten08, gunawardhana11}.  
This suggests that commonly adopted model assumptions
regarding dust attenuation or the initial mass function may be too
simplistic.  For instance, there is some evidence for a dust component associated with 
intermediate age stars (Eminian et al. 2008).  Dust attenuation is also likely to be inhomogeneous 
within a given galaxy, and the net effect may not be well approximated by our two `effective' global 
parameters, $\mu$ and $\tau_V$.  The importance of dust inhomogeneities has been demonstrated 
in a study of 9 local galaxies using multiband photometry \citep{zibetti09}.  Further 
exploration of the differences between spatially resolved and unresolved SFR and mass estimates 
will be possible with the next generation of integral field unit galaxy surveys \citep[c.f.,][]{sanchez11}.  
We defer a full analysis of the
difference between PCA and nebular estimates of the SFR to future
work. In $\S7$ we will compare PCA-derived $F*$ values for DR7 and
BOSS galaxies at fixed stellar mass.  While the absolute values of
$F*$ are somewhat uncertain, our analysis hinges on the relative
differences which we believe to be robust.

\bibliographystyle{mn2e}

\bibliography{mass}

\end{document}